\documentclass[aps,prd,preprint]{revtex4}

\usepackage[dvips]{graphicx}
\usepackage{amssymb}
\usepackage{amstext}
\usepackage{amsmath}

\def\lp {\left( }
\def\rp {\right) }
\def\lb {\left[ }
\def\rb {\right] }
\def\lc {\left\{ }
\def\rc {\right\} }
\def\ra {\rangle }
\def\la {\langle }

\def\rar {\rightarrow}

\def\beq{\begin{equation}}
\def\eeq{\end{equation}}
\def\bea{\begin{eqnarray}}
\def\eea{\end{eqnarray}}
\def\nn {\nonumber}
\def\ni {\noindent}

\def\dr {\partial}

\def\cO {{\cal O}}

\def\a{\alpha}
\def\b{\beta}
\def\d{\delta}
\def\e{\epsilon}

\def\l{\lambda}

\def\m{\mu}

\def\p {\pi}

\def\s{\sigma}

\def\th {\theta}

\def\bfi {\mbox{\boldmath $\phi$}}
\def\bb {\mbox{\boldmath $b$}}

\def\bP {\mbox{\boldmath $P$}}
\def\bQ {\mbox{\boldmath $Q$}}

\def\bu {\mbox{\boldmath $u$}}
\def\bv {\mbox{\boldmath $v$}}

\def\btau {\mbox{\boldmath $\tau$}}

\def\bl {\mbox{\boldmath $\l $}}

\def\bpi {\mbox{\boldmath $\pi$}}
\def\bphi {\mbox{\boldmath $\phi$}}
\def\btau {\mbox{\boldmath $\tau$}}

\begin{document}

%\maketitle{analytic $SU(3) $ unitary matrix  and application to a $\s$-model}

\title{chiral symmetry: an analytic $SU(3) $ unitary matrix}

\author{M.R. Robilotta}
\affiliation{Instituto de F\'isica,  Universidade de S\~ao Paulo, S\~ao Paulo, Brazil}

\date{\today }

\begin{abstract}

The $SU(2)$ unitary matrix $U$ employed in chiral descriptions of hadronic 
low-energy processes   
has both exponential and analytic representations, related by
 $ U = \exp\lb \, i\, \btau \cdot \hat{\bpi}  \, \theta\,\rb 
=  \cos\th \,I + i\, \btau \cdot \hat{\bpi}  \, \sin\th $, 
where $\btau$ are Pauli matrices and   $\bpi = (\p_1, \p_2,\p_3) $
is the pion field.
One extends  this result  to the $SU(3)$ unitary matrix
by deriving an analytic expression which, for Gell-Mann matrices $\bl$, reads
$ U= \exp\lb\,  i\, \bv\cdot \bl \, \rb
=  \lb \,  \lp F + \tfrac{2}{3}\, G \rp I 
+ \lp H \, \hat{\bv} + \tfrac{1}{\sqrt{3}}\,G \, \hat{\bb} \rp \cdot \bl \, \rb
+\,  i\, \lb \,  \lp Y + \tfrac{2}{3}\, Z \rp I 
+ \lp X \, \hat{\bv}  +  \tfrac{1}{\sqrt{3}}\,Z \, \hat{\bb} \rp \cdot \bl \, \rb $,
with $v_i=[\,v_1, \cdots  v_8\,]$, $ b_i = d_{ijk} \, v_j \, v_k $,
and factors $F, \cdots Z$ written in terms of elementary functions 
depending on $v=|\bv|$ and 
$\eta = 2\, d_{ijk} \, \hat{v}_i \, \hat{v}_j \, \hat{v}_k /3 $.
This result does not depend on the particular meaning attached to the variable $\bv$
and the analytic expression  is used to calculate explicitly 
the associated left and right forms.
When $\bv$ represents pseudoscalar meson fields,
the classical limit corresponds to $\la 0|\eta|0\ra \rar \eta \rar 0$ and
yields the cyclic structure
$ U = \lc \lb \tfrac{1}{3} \lp 1 + 2\, \cos v \rp \, I
+ \tfrac{1}{\sqrt{3}} \lp -1 + \cos v  \rp \, \hat{\bb}\cdot \bl \rb 
+ i\, \lp \sin v \rp \hat{\bv}\cdot \bl \rc $,
which gives rise to a tilted circumference with radius $\sqrt{2/3}$
in the space defined by $I$, $\hat{\bb}\cdot \bl $, and $\hat{\bv}\cdot \bl $.
For the sake of completeness, the axial transformations of the analytic matrix are also evaluated explicitly.

\end{abstract}

\pacs{...}

\maketitle

\section{motivation}
\label{mot}

The considerable progress in low-energy hadron physics achieved
over the last sixty years is closely associated with chiral symmetry. 
Quantum chromodynamics (QCD), the present-day strong theory, involves gluons and
six quarks with different flavors, which have color.
Direct applications to low-energy processes are very difficult owing to 
gluon–gluon interactions 
and one has to resort to either lattice methods\cite{LattQCD} or effective descriptions.
The latter depart from the symmetries of QCD, 
namely the  continuous Poincar\'e group, discrete C, P and T inversions, 
together electric charge and baryon number conservation.
The quark masses $m_q$ are external parameters and the lightest ones,
$m_u$, $m_d$, and $m_s$, can be considered as small in the scale 
$\Lambda \sim 1\, \mathrm{GeV}$. 
This rationale underlies the idea of chiral symmetry, 
an approximate scheme that becomes exact in the 
ideal limit $m_q\rar 0$. 
In this case, helicity is a good quantum number and 
the quark fields $q$ are written as
linear combinations of $q_R$ and $q_L$, with spins respectively 
parallel and anti-parallel to their momenta. 
As helicity is conserved in interactions, the fields $q_R$ and $q_L$ do not couple and the  
Lagrangian is symmetric under the chiral group $U(N)_R \times U(N)_L$, 
where N is the number of flavors.
However, owing to the the $U(1)_A$ anomaly, the actual group to be considered is 
$U(1)_V \times SU(3)_R \times SU(3)_L $.
In effective descriptions, these symmetries of QCD are associated directly with 
hadronic degrees of freedom, bypassing quarks and gluons.

The incorporation of chiral symmetry into hadron physics precedes QCD
and was already being discussed in 1960.
A long-lasting contribution from that year is the idea that the strong vacuum
is not empty, presented by Gell-Mann and L\'evy\cite{GellMannLevy} 
in a paper introducing both linear 
and non-linear $\sigma$-models for pions and nucleons.
The former  relied on the $\s$, a scalar particle proposed 
earlier by Schwinger\cite{Schwinger}, and 
provided a unique tool for  dealing with the strong vacuum.
In the symmetric version, the model involves just two parameters, 
usually denoted by $\m$ and $\l$, whose values determine whether 
the ground state of the theory is either empty or contains  a classical component,
associated with a condensate.
Almost simultaneously, in 1961, Nambu and Jona-Lasinio\cite{NambuJonaL} 
studied the strong vacuum employing an alternative chiral model 
inspired by superconductivity,  which also involved a scalar-isoscalar state.
Their model was based on fermionic fields, the pion being a collective state, 
and contained  a vacuum phase transition 
described by a gap-equation, controlled by a free parameter.
A common feature of both models is the indication that chiral symmetry allows
the ground state of strong systems to be realized in two different ways, namely:
(i) the Wigner–Weyl mode, in which states with opposite parities are degenerate and
the vacuum is empty;
(ii) the Nambu–Goldstone mode, in which the pion is a massless Goldstone boson,
the scalar state is massive  and  the vacuum contains a condensate.
Also in 1961, Skyrme succeeded in describing baryons as 
topological solitons composed of chiral pions, 
carrying a well defined quantum number\cite{Skyrme}.
He employed classical pion fields  constrained by a non-linear condition 
and assumed the proton to be a deformation of the strong vacuum,
kept stable for topological reasons. 
Nowadays, these states are known as skyrmions but, at the time, 
they were criticized for not having spin and  deserved little attention.
However, about two decades later, spins were incorporated into the model by 
Adkins, Nappi and Witten\cite{AdNaWit}, and its rich structure could be properly appreciated.

After QCD became established as the strong theory, applications of chiral symmetry 
were aimed mostly at improving the precision of predictions and
nowadays chiral perturbation theory (ChPT) is employed to tackle low-energy 
hadronic processes.
This research program was outlined by Weinberg in 1979\cite{Weinberg1} and
fully developed by Gasser and Leutwyler for the SU(2) sector in 1984\cite{GassLeut1}.
Low-energy interactions are strongly dominated by  quarks $u$ and $d$ and
their small masses are treated as  perturbations
into a massless $SU(2) \times SU(2)$ symmetric Lagrangian
involving effective pion fields.
ChPT is a well-defined theory and allows the systematic expansion 
of low-energy amplitudes in powers of a typical scale $q \sim M_\p< 1 \,\mathrm{GeV}$. 
Nevertheless, while QCD is fully renormalizable, ChPT can only be renormalized order by 
order\cite{Weinberg1}.
The effective lagrangian consists of strings of terms
possessing the most general structure consistent with broken chiral symmetry
and both its form and  the number of low-energy constants (LECs) 
associated with renormalization  depend on the order considered.

All approaches to strong interactions mentioned, 
namely the models produced by Gell-Mann and L\'evy, Nambu and Jona-Lasinio,
and Skyrme, together with ChPT, did bring important progress to the area.
With hindsight, however, one realizes that all of them have specific limitations
and none has superseded completely the others. 
So, in spite of their differences, they coexist and the relevance of each one
depends on the particular problem considered.
A common feature of these competing strategies is that,
in all cases, early works were performed in the framework of $SU(2)$
for reasons of simplicity.
The basic unitary $SU(2)$ matrix $U$ can be represented as
\bea 
&& U= \exp\lb \, i\, \btau \cdot \hat{\bpi}  \, \theta\,\rb \;,
\hspace{5mm} \rar \;\; \mathrm{exponential} \; \mathrm{representation}
\label{mot.1}
\eea 
where $\hat{\bpi}$ is the direction of the pion field in isospin space
and $\theta$ is the chiral angle.
As it is well known, the series implicit in the exponential can be summed 
and one gets  the equivalent  form 
\bea 
&& U= \cos\th\,I + i\, \btau \cdot \hat{\bpi}  \, \sin\th  \;,
\hspace{5mm} \rar \;\;\mathrm{analytic} \; \mathrm{representation}
\label{mot.2}
\eea 
which one calls analytic, in the want of a better name.
It is employed in the non-linear $\s$-model and suited to comparisons 
with the linear version, based on the non-unitary matrix
\bea
&& M = \s\,I + i\, \btau \cdot \bpi  \;.
\label{mot.3}
\eea 
The simplicity of these structures facilitates 
comparisons among different schemes 
and allows one to study the mathematical reasons
behind their main features.

The various approaches have been generalized to $SU(3)$ and
this version of the $\s$-model\cite{Levy}
employs a matrix $M$ composed by nonets of pseudoscalar and scalar states,
whereas the extended version of ChPT relies on the exponential form\cite{GassLeut2}.
In the case of the Skyrme model,
the $SU(3)$ group is employed just in the  quantization of the soliton,
which is carried out formally\cite{Chemtob,ChemtobBlaizot}. 
The conceptual mobility among these generalizations to $SU(3)$ is more difficult 
than in $SU(2)$, partly owing to the absence of a suitable analytic 
expression for the matrix $U$
which could provide a bridge among them.
Analytic results based on Euler angles 
already exist for this matrix \cite{Nelson, AkyRas, Byrd}
and find applications in many areas of physics dealing
with three state systems, such as color superconductivity\cite{color}, optics\cite{optics}, 
geometric phases\cite{geompha},
and quantum entanglement in computation and communication\cite{entanglement}.
However,  Euler angles require a set of external axes and are inconvenient
to applications of chiral symmetry to low-energy processes.
In sect.\ref{s2} one derives an alternative analytic representation for the matrix $U$,
written in terms of internal degrees of freedom and
corresponding to an extension of eq.(\ref{mot.2}).
In hadron physics, this result may be
instrumental to simplifying calculations and 
studying topological properties of $SU(3)$, for both flavor and color,
in analogy to the case of the skyrmion.
The unitarity of $U$ in analytic form is explored in sect.\ref{s3}
and the corresponding left and right forms are presented in sect.\ref{s5}.
Its classical limit is discussed in sect.\ref{s4},
chiral transformations of are given in sect.\ref{s6},
conclusions are summarized in sect.\ref{summa}, and
technical matters are presented in four appendices.

%-------------------------------------------------------------------------------
 \section{analytic form}
\label{s2}

The exponential form of $U$ in $SU(3)$ is
written in terms of the Gell-Mann matrices $\bl = [\, \l_1, \cdots \l_8 \,] $,
satisfying $ \lb \l_i , \l_j \rb = 2\, i\, f_{ijk}\,\l_k $ and 
$ \lc \l_i , \l_j \rc = \tfrac{4}{3}\, \d_{ij}\,I + 2\, d_{ijk}\,\l_k $,
coupled with  a generic octet  $\bv=[\,v_1, \cdots  v_8\,]$  as
\bea 
&& U = \exp\lb\,  i\, \bv\cdot \bl \, \rb 
\nn\\[2mm]
&& \hspace{5mm} = \lb 1 - \frac{v^2}{2!} \lb \hat{\bv} \cdot \bl \rb^2 + \cdots  \rb 
+ i \,\lb \frac{v}{1!} \lb \hat{\bv} \cdot \bl \rb  - \frac{v^3}{3!} \lb \hat{\bv} \cdot \bl \rb^3
+ \cdots  \rb
\label{anf.1}
\eea
with $ \bv\cdot \bl = v_i\, \l_i $, $v = \sqrt{ v_i \,  v_i}  $, and $\hat{\bv}=\bv/v$.

One uses two auxiliary variables in the derivation of the analytic form.
One of them is the bilinear construct $\bb=[\,b_1, \cdots  b_8\,]$,
\bea
b_i = d_{ijk} \, v_j \, v_k \;,
\label{anf.2}
\eea
with $b= \sqrt{\,b_i\, b_i} = v^2/\sqrt{3} $ and $\hat{\bb}=\bb/b$.
The other is
\bea
&& \eta = \frac{2}{3\,v^3} \; D = \frac{2}{3\,\sqrt{3}}\; \hat{\bv}\cdot \hat{\bb} \;,
\label{anf.3}
\eea
where $D=\bv \cdot \bb = d_{ijk} \, v_i \, v_j \, v_k  $.
The quantity $\bb $ is even under $\bv \rar -\bv$, $\eta $ is odd, and the latter
is a measure of the overlap between $\bb $ and $\bv$.
As $\hat{\bv}$ and $\hat{\bb}$  are unit vectors, $|\hat{\bv}\cdot \hat{\bb}|\leq 1$,
and $\eta$ lies in the interval $- \tfrac{2}{3\,\sqrt{3}} \leq \eta \leq \tfrac{2}{3\,\sqrt{3}}$.
The explicit forms of $b_i$ and $D$ are given in App.\ref{auxfunc} and shown to
satisfy the conditions
\bea
&& f_{ijk} \, v_j \, b_k = 0\;,
\label{anf.4}\\[2mm]
&& d_{ijk} \, v_j \, b_k = \tfrac{1}{3}\, v^2 \, v_i \;,
 \label{anf.5}\\[2mm]
&& d_{ijk} \, b_j \, b_k = \eta\, v^3 \, v_i - \tfrac{1}{3}\, v^2 \, b_i \;,
 \label{anf.6}
\eea
which allow one to write
\bea
&& \lb\, \hat{\bv}\cdot \bl \,\rb \, \lb \hat{\bv}\cdot \bl \, \rb 
= \tfrac{2}{3}\, I  +\tfrac{1}{\sqrt{3}}   \;  \hat{\bb} \cdot \bl  \;, 
\label{anf.7}
\\[2mm]
&& \lb\, \hat{\bv}\cdot \bl \,\rb \, \lb \hat{\bv}\cdot \bl \, \rb \, \lb \hat{\bv}\cdot \bl \, \rb 
=  \hat{\bv} \cdot \bl  + \eta \,I\;.
\label{anf.8}
\eea
In order to simplify the notation, one defines
\bea 
&& A =  \hat{\bv}\cdot \bl \;,
\label{anf.9}\\[2mm]
&& B = \tfrac{2}{3}\,I  +\tfrac{1}{\sqrt{3}}\; \hat{\bb} \cdot \bl \;,
\label{anf.10}
\eea
so that 
\bea 
&& A^2=  A A = B \;,
\label{anf.11}\\[2mm]
&& A^3= AB = A + \eta\,I \;,
\label{anf.12}\\[2mm]
&& A^4= AAB = BB = B + \eta \, A \;.
\label{anf.13}
\eea
Thus, $ A^5 = A \lp B +\eta\,A \rp  = A + \eta \,I+ \eta\, B $, and so on.
These results mean that, in matrix space, $U$ is a 
linear combinations of $I$, $A$ and $B$.
The matrices $I$ and $B$ are even under $ \bv \rar - \bv $, 
whereas $A$ is odd.

The matrix $U$ is written as
\bea
U =\sum_{n=0}^\infty 
 i^n \, \frac{v^{n}}{{n}!}\, A^n 
 = U_\mathrm{even} + i\, U_\mathrm{odd} \;,
\label{anf.14}
\eea
where the labels {\it even} and {\it odd} refer to $ \bv \rar - \bv $,
and one has
\bea
&& \frac{\dr U_\text{even}}{\dr v}  = - A\, U_\text{odd} \;,
\label{anf.15}\\[2mm]
&& \frac{\dr U_\text{odd}}{\dr v}  = A\, U_\text{even} \;.
\label{anf.16}
\eea
In matrix space, one writes
\bea
&& U_\text{even} = \sum_{n=0}^\infty 
 (i)^{2n} \frac{v^{2n}}{(2n) !} \lb\,  f_{2n} \,I + g_{2n} \,B + h_{2n} \,A \,\rb 
\nn\\[2mm]
&& \hspace{6mm} =  
 F(v, \eta)\,I + G(v, \eta) \,B + H(v, \eta) \,A \;,
\label{anf.17}\\[2mm]
&& U_\text{odd}  = \sum_{n=0}^\infty  (i)^{2n}  \frac{v^{2n+1}}{(2n+1)!} 
\lb \, x_{2n+1} \,A + y_{2n+1}\,I + z_{2n+1} \,B \,\rb
\nn\\[2mm]
&& \hspace{7mm}=   X(v, \eta)\,A + Y(v, \eta) \,I + Z(v, \eta) \,B \;,
\label{anf.18}
\eea
where the functions $F, G, H, X, Y, \mathrm{and}\, Z$ are determined in the sequence.
In tables \ref{T1} and \ref{T2}, one displays a few partial contributions
to these series and it is possible
to note that the dependences on $v$  and $\eta $ do not mix.
Even and odd components are related by the action of the matrix A, 
which yields
\bea
&& A\,\lb \, f_{2n} \, I + g_{2n}\, B + h_{2n} \,A \,\rb 
= \lb \, (f_{2n} + g_{2n}) \, A + \eta \,  g_{2n} \,I + h_{2n} \, B \,\rb
\nn\\[2mm]
&& \hspace {5mm}
= \lb \, x_{2n+1} \, A + y_{2n+1} \,I + z_{2n+1} \,B \,\rb \;,
\label{anf.19}\\[2mm]
&& A\,\lb \, x_{2n+1} \, A + y_{2n+1}\, I + z_{2n+1} \,B \,\rb 
= \lb \,  \eta\,  z_{2n+1} \,I  +  x_{2n+1} \, B   + ( y_{2n+1} + z_{2n+1} ) \,A  \,\rb
\nn\\[2mm]
&& \hspace {5mm}
= \lb \, f_{2n+2} \, I + g_{2n+2} \,B + h_{2n+2} \,A \,\rb \;.
\label{anf.20}
\eea

\renewcommand{\arraystretch}{1.4}
\begin{table}[!h]
\begin{center}
\begin{tabular}{c c l l l}
\hline
n &\hspace{4mm} $ (i\,v)^n/n! $ \hspace{8mm} & $f_n \,(\times I) $ \hspace{10mm} 
&$g_n\, (\times B) $\hspace{18mm} 
& $ h_n\, (\times A)$ \hspace{10mm} \\ \hline
  
0 &  1    & 1  &   &                      \\ \hline

2 &  $-v^2/2!$    &   & 1  &                                                      \\ \hline

4 &  $v^4/4!$    &   & 1  & $\eta$                            \\ \hline

6 &  $-v^6/6!$    & $\eta^2$  & 1  &  $2\eta$                 \\ \hline

8 &  $v^8/8!$    & $2\,\eta^2$  &  $1+\eta^2$ & $3\,\eta$    \\ \hline

10 &  $-v^{10}/10!$    & $3\,\eta^2$  & $1+ 3\,\eta^2$  &  $4\,\eta + \eta^3$      \\ \hline

12 &  $v^{12}/12!$    & $4\,\eta^2 + \eta^4$  & $1+ 6\,\eta^2$   & $5\,\eta + 4\,\eta^3$ \\ \hline

14 &  $-v^{14}/14!$    & $5\,\eta^2 + 4\, \eta^4$  & $1 + 10\,\eta^2 + \eta^4$  
&  $6\,\eta + 10\, \eta^3$ \\ \hline

16 &  $v^{16}/16!$    & $6\,\eta^2 +10\,\eta^4$  &  $1 + 15\,\eta^2 + 5\,\eta^4$
&  $7\,\eta + 20\, \eta^3 + \eta^5 $ \\ \hline

\hline
\end{tabular}
\end{center}
\caption{Structure of $U_\text{even} $, eq.(\ref{anf.17}). }
\label{T1}
\end{table}

\renewcommand{\arraystretch}{1.4}
\begin{table}[!h]
\begin{center}
\begin{tabular}{c c l l l}
\hline
n & \hspace{4mm} $ (i\,v)^n/n! $ \hspace{8mm} & $x_n\,(\times A)$ \hspace{12mm} 
&$y_n\,(\times I) $\hspace{18mm} 
& $z_n \, (\times B)$ \hspace{10mm} \\ \hline

1 &  $i\,v$    &  1  &   &                       \\ \hline

3 &  $-i\,v^3/3!$    & 1  & $\eta$  &     \\ \hline

5 &  $i\,v^5/5!$    & 1  & $\eta$  & $\eta$      \\ \hline

7 &  $-i\,v^7/7!$    & $1+\eta^2$   & $\eta$  & $2\,\eta$   \\ \hline

9 &  $i\,v^9/9!$    & $1+ 3\, \eta^2$   & $\eta + \eta^3$  & $3\,\eta$  \\ \hline

11 &  $-i\,v^{11}/11!$    & $1+6\,\eta^2$  & $\eta+ 3\,\eta^3$  & $4\,\eta+\eta^3$
 \\ \hline

13 &  $i\,v^{13}/13!$    & $1 + 10\,\eta^2 + \eta^4$  & $\eta + 6\,\eta^3$  
& $5\,\eta +4\, \eta^3$  \\ \hline

15 &  $-i\,v^{15}/15!$    & $1+ 15\,\eta^2 + 5\,\eta^4$  &  $\eta + 10\,\eta^3 + \eta^5$ 
&  $6\,\eta + 10\,\eta^3$ \\ \hline

17 &  $i\,v^{17}/17!$    & $1+21\,\eta^2 + 15\,\eta^4$  & $\eta+15\,\eta^3 +5\,\eta^5$  
& $7\,\eta+20\,\eta^3+\eta^5$ \\ \hline

\hline
\end{tabular}
\end{center}
\caption{ Structure of $i\, U_\text{odd} $, eq.(\ref{anf.18}).}
\label{T2}
\end{table}

Using results (\ref{anf.15})-(\ref{anf.18}), one writes
\bea
&& \frac{\dr}{\dr v} \lb F\, I + G\, B + H\, A \rb 
= - \lb \eta\,Z\,I + X\,B + (Y+Z) A \rb \;,
\label{anf.21}\\[2mm]
&& \frac{\dr}{\dr v}  \lb X\, A + Y\, I + Z\, B \rb
= \lb \lp  F+G \rp A + \eta\, I + H\,B \rb \;.
\label{anf.22}
\eea
and obtains a set of first order differential equations
coupling even and odd components
\bea
&& \frac{\dr F}{\dr v}  = - \eta\, Z\;,
\hspace{12mm}
\frac{\dr G}{\dr v}  = - X\;,
\hspace{12mm}
\frac{\dr H}{\dr v}  = -Y - Z \;,
\label{anf.23}\\[2mm]
&& \frac{\dr X}{\dr v}  = F+G \;,
\hspace{10mm}
\frac{\dr Y}{\dr v}  =  \eta\, G \;,
\hspace{13mm}
\frac{\dr Z}{\dr v}  = H  \;.
\label{anf.24}
\eea 
A further derivation decouples these sectors, yielding
\bea
&& \frac{\dr^2 F}{\dr v^2}  = -\eta \, H  \;,
\hspace{16mm}
\frac{\dr^2 G}{\dr v^2}  = - F - G \;,
\hspace{10mm}
\frac{\dr ^2 H}{\dr v^2}  = - \eta\, G - H \;,
\label{anf.25}\\[2mm]
&& \frac{\dr^2 X}{\dr v^2}  = - \eta\, Z - X \;,
\hspace{8mm}
\frac{\dr^2 Y}{\dr v^2}  =  -\eta \, X \;,
\hspace{15mm}
\frac{\dr^2 Z}{\dr v^2}  = -Y - Z \;. 
\label{anf.26}
\eea 
In order to get an uncoupled differential equation for $F$, one increases 
the number of derivatives and finds
\bea
&& \frac{\dr^6F}{\dr v^6}  + 2\, \frac{\dr^4F}{\dr v^4}  + \frac{\dr^2F}{\dr v^2}   
+ \eta^2 \,F  =0 \;.
\label{anf.27}
\eea
Its general solution is discussed in App.\ref{cubic} and given by
\bea
F = \b_1 \, \cos(k_1\,v) + \b_2 \, \cos(k_2\,v) + \b_3 \, \cos(k_3\,v) \;,
\label{anf.28}
\eea 
where $\b_i$ are constants and, using the results of App.\ref{cubic}, one has
\bea
&& k_i = \tfrac{2}{\sqrt{3}}\, \sin (\th/6 + \d_i \, \pi/3) \;,
\label{anf.29}\\[2mm]
&& \cos(\theta) = 1 - 27\,\eta^2/2\;,
\hspace{5mm}
\sin(\theta) = 3\sqrt{3}\, \eta \, \sqrt{1-27\,\eta^2/4} \;,
\label{anf.30}
\eea
and $\d_1=0, \d_2=1, \d_3=-1$.

The constants $\b_i$ are fixed by expanding $\cos (k_i\,v)$ in series and,
expressing results in terms of the  roots $\a_i=-k_i^2$ of the cubic equation
$ \a_i^3 + 2\, \a_i^2 + \a_i + \eta^2 =0 $, eq.(\ref{B.3}),
one has
\bea
&& F = (\b_1 + \b_2 + \b_3) + \frac{v^2}{2!} \,(\b_1\,\a_1 + \b_2 \, \a_2 +\b_3\, \a_3)
+ \frac{v^4}{4!}\,(\b_1\, \a_1^2 + \b_2 \, \a_2^2 +\b_3\, \a_3^2)
\nn\\[2mm]
&& \hspace{5mm} + \frac{v^6}{6!} \,\sum \b_i\, \a_i^3 
+ \frac{v^8}{8!} \, \sum  \b_i\, \a_i^4 
+\, \frac{v^{10}}{10!} \, \sum  \b_i\, \a_i^5 
+ \frac{v^{12}}{12!} \, \sum  \b_i\, \a_i^6 + \cdots \;.
\label{anf.31}
\eea
Comparing results for $v^0$, $v^2$ and $v^4$ with those of table \ref{T1}, 
one learns that
\bea
&& \b_1 + \b_2 + \b_ 3 = 1 \;,
\label{anf.32}\\[2mm]
&& \b_1\,\a_1 + \b_2 \, \a_2 +\b_3\, \a_3 =0 \;,
\label{anf.33}\\[2mm]
&& \b_1\, \a_1^2 + \b_2 \, \a_2^2 +\b_3\, \a_3^2 = 0 \;.
\label{anf.34}
\eea
Terms proportional to powers of $v$ $\geq 6$ are evaluated using combinations 
of eqs.(\ref{anf.32})-(\ref{anf.34}) and (\ref{B.3}).
Thus, for instance 
\bea
&& \sum \b_i\, \a_i^3 
= \sum \, \b_i  \lb - 2\, \a_i^2 - \a_i - \eta^2 \rb = -\eta^2 \;,
\label{anf.35}\\[2mm]
&& \sum \b_i\, \a_i^4 
= \sum \, \b_i \, \a_i \, \lb - 2\, \a_i^2 - \a_i - \eta^2 \rb 
= 2\,\eta^2 \;,
\label{anf.36}\\[2mm]
&& \sum \b_i\, \a_i^5  
= \sum \, \b_i \, \a_i^2 \, \lb - 2\, \a_i^2 - \a_i - \eta^2 \rb 
= - 3\,\eta^2 \;,
\label{anf.37}\\[2mm]
&& \sum \b_i\, \a_i^6 
= \sum \, \b_i \,  \lb - 2\, \a_i^2 - \a_i - \eta^2 \rb^2 
= 4\,\eta^2 + \eta^4 \;.
\label{anf.38}
\eea
Using these results into eq.(\ref{anf.28}), one has
\bea
F =1  - \frac{v^6}{6!} \,\eta^2 + \frac{v^8}{8!} \,2\,\eta^2 
- \frac{v^{10}}{10!} \, 3\, \eta^2 + \frac{v^{12}}{12!} \,\lp 4\,\eta^2 + \eta^4 \rp + \cdots
\label{anf.39}
\eea
and the entry in table \ref{T1}  is reproduced.

Expressions (\ref{anf.32})-(\ref{anf.34}) yield directly
\bea
&& \b_i = -\, \frac{\a_j \, \a_k}{(\a_i - \a_j) \, (\a_k - \a_i )} \;,
\label{anf.40}
\eea
with $ [i, j, k] \rar \mathrm{cyclic}\; \mathrm{permutations} \; \mathrm{of} \; [1, 2, 3]$.
Alternative versions are useful in calculations and,
employing condition (\ref{B.3}), on has 
\bea
&& \b_i =  \frac{\eta^2 }{\a_i \,(\a_i - \a_j) \, (\a_k - \a_i )} \;.
\label{anf.41}
\eea
The denominator can be simplified using results of App.\ref{cubic} and 
one finds the set of alternatives
\bea
&& \b_i = \frac{\eta^2}{2\,(\a_i^2+ \a_i) + 3\, \eta^2} \;,
\label{anf.42}\\[2mm]
&& \b_i = - \,\frac{\eta^2}{(\a_i^2+ \a_i)\,(3\,\a_i+1)} \;,
\label{anf.43}\\[2mm]
&& \b_i = \frac{\a_i+1}{(3\,\a_i+1)} \,.
\label{anf.44}
\eea
The last expression determines the condition
\bea
&& 
\sum_i \frac{1}{(3\,\a_i+1)}  = 0 \;.
\label{anf.45}
\eea

Result (\ref{anf.28}) for $F$  and eqs.(\ref{anf.23})-(\ref{anf.26}) determine
the set of functions $G$, $H$, $X$, $Y$, and $Z$.
Choosing form (\ref{anf.44}) for the $\b_i$, one has
\bea
&& F = \frac{(\a_1+1)}{ (3\,\a_1 +1)} \; \cos(k_1\,v) + [1\rar 2, 3] \;,
\label{anf.46}\\[2mm]
&&G = -\, \frac{1}{(3\,\a_1 +1)} \; \cos(k_1\,v) +  [1\rar 2, 3] \;,
\label{anf.47}\\[2mm]
&&H = \frac{\eta }{(\a_1+1)\, (3\,\a_1+1)} \; \cos(k_1\,v) +  [1\rar 2, 3] \;,
\label{anf.48}\\[2mm]
&& X = -\, \frac{1}{(3\,\a_1 +1)} \; k_1 \sin(k_1\,v) +  [1\rar 2, 3] \;,
\label{anf.49}\\[2mm]
&& Y = \frac{\eta}{\a_1 \, (3\,\a_1+1)} \; k_1 \sin(k_1\,v) +  [1\rar 2, 3] \;,
\label{anf.50}\\[2mm]
&& Z = -\, \frac{\eta}{\a_1\,(\a_1+1)\,(3\,\a_1+1)} \; k_1 \sin(k_1\,v) +  [1\rar 2, 3] \;.
\label{anf.51}
\eea
The parity of these functions under $\bv \rar -\bv$ is determined by 
$\eta$ and therefore $F$, $G$, and $X$ are even, 
whereas $H $, $Y $, and $Z $ are odd.

The analytic form of the matrix $U$, derived from eqs.(\ref{anf.17}) 
and (\ref{anf.18}),  reads
\bea
&& U= \lp F\,I + G \, B + H \, A \rp + i \lp Y \,I + Z\, B +  X\, A \rp \;,
\label{anf.52} \\[2mm]
&& \hspace{4mm} =  \lb \,  \lp F + \tfrac{2}{3}\, G \rp \,I 
+ \lp H \, \hat{\bv} + \tfrac{1}{\sqrt{3}}\,G \, \hat{\bb} \rp \cdot \bl \, \rb
\nn\\[2mm]
&& \hspace{4mm}
+\,  i\, \lb \,  \lp Y + \tfrac{2}{3}\, Z \rp \,I 
+ \lp X \, \hat{\bv}  +  \tfrac{1}{\sqrt{3}}\,Z \, \hat{\bb} \rp \cdot \bl \, \rb \;.
\label{anf.53}
\eea
As there is an overlap between $\hat{\bv}$  and $\hat{\bb}$, one might consider
replacing the latter by the unit vector $\hat{\bu}$ 
given by
\bea
\hat{\bb}= \frac{3\,\sqrt{3}}{2}\,\eta\, \hat{\bv} + \sqrt{1-\frac{27\,\eta^2}{4}}\, \hat{\bu} \;,
\label{anf.54}
\eea
such that $\hat{\bv}\cdot\hat{\bu}=0 $.
However, this is not especially useful.

In order to deal with a more compact expression, one defines the quantities
\bea
&& S = \lp F + \tfrac{2}{3}\,G \rp \;,
\hspace{15mm}
Q_i = \sqrt{\tfrac{2}{3}}\, \lp H\, \hat{v}_i +\tfrac{1}{ \sqrt{3}}\,G\, \hat{b}_i \rp \;,
\label{anf.55}\\[2mm]
&& W = \lp Y + \tfrac{2}{3}\,Z \rp \;,
\hspace{15mm}
P_i =  \sqrt{\tfrac{2}{3}}\, \lp  X\, \hat{v}_i + \tfrac{1}{\sqrt{3}}\,Z\, \hat{b}_i  \rp \;,
\label{anf.56}
\eea
and expresses the equivalence between exponential and analytic representations as
\bea
&& U = \exp\lb\,  i\, \bv\cdot \bl \, \rb 
=
\lb \, S \,I  + \sqrt{\tfrac{3}{2}} \, \bQ \cdot \bl \, \rb
+ i\, \lb \, W \,I  + \sqrt{\tfrac{3}{2}} \, \bP \cdot \bl \, \rb \;.
\label{anf.57}
\eea
%

%--------------------------------------------------------------------------------
\section{unitarity}
\label{s3}

The matrix $U$ satisfies the $SU(3)$ conditions $U^\dagger \, U=I$ 
and  $\text{det} \, U=1$ irrespective of the representation adopted, 
as ensured by result (\ref{anf.57}).
Nevertheless, it is useful to explore the unitarity condition expressed in 
analytic form given by eqs.(\ref{anf.52}) and (\ref{anf.53}),
since it gives rise to constraints among the factors $F, \cdots Z$.
Explicit multiplication using form (\ref{anf.52}), together with eqs.(\ref{anf.11})-(\ref{anf.13}),
yields
\bea
U^\dagger \,U = C_I \, I + C_B \, B + C_A \, A  \;,
\label{unit.1}
\eea
with
\bea
&& C_I = F^2 + 2\,\eta \, G H + Y^2 + 2\,\eta\,X Z \;,
\label{unit.2}\\[4mm]
&& C_B = G^2 + H^2 + 2\, F G + X^2 + Z^2 + 2\,Y Z \;,
\label{unit.3}\\[4mm]
&& C_A = \eta\, G^2 + 2\, F H + 2\, G H + \eta\, Z^2 + 2\, X Y + 2\, X Z \;,
\label{unit.4}
\eea
and, in  App.\ref{unitproof}, one shows that 
\bea
&& C_I = 1\;, \hspace{10mm} C_B=0\;, \hspace{10mm} C_A = 0\;.
\label{unit.5}
\eea

Alternatively, form (\ref{anf.57}) gives rise to 
\bea
&& U^\dagger \, U= \lb S^2 +  Q^2 + W^2 + P^2\,\rb I
\nn\\[2mm]
&& \hspace{5mm} + \lc \sqrt{6}\,S  Q_k + \tfrac{3}{2} \,Q_i Q_j\, d_{ijk}
+  \sqrt{6}\,W P_k + \tfrac{3}{2}\, P_i P_j\, d_{ijk} 
+ \tfrac{3}{2}\, Q_i P_j\, f_{ijk} \rc \l_k \;.
\label{unit.6}
\eea
Definitions (\ref{anf.55}), (\ref{anf.56}), with
results (\ref{anf.2}), (\ref{anf.5})  and (\ref{anf.6}), allow one to show that
the term within curly brackets is
\bea
&& \lc  \cdots \rc =  C_A \, \hat{v}_k + \tfrac{1}{\sqrt{3}}\, C_B\, \hat{b}_k  = 0 \;.
\label{unit.7}
\eea
Writing 
\bea
&& Q^2=  Q_i \, Q_i =\tfrac{2}{9}\, G^2 +\tfrac{2}{3}\, H^2 + 2\, \eta \,G H \;,
\label{unit.9a}\\[2mm]
&& P^2  = P_i\,P_i = \tfrac{2}{3}\, X^2 +\tfrac{2}{9}\, Z^2 \, + 2\, \eta \, X Z \;,
\label{unit.9b}
\eea
one also has
\bea 
&& U^\dagger \, U=  \lb S^2 +  Q^2 + W^2 +  P^2 \, \rb I 
\nn\\[2mm]
&& \hspace{5mm} = \lb F^2 + \tfrac{4}{3}\, F G + \tfrac{2}{3}\, G^2 
+ \tfrac{2}{3}\, H^2 + 2\, \eta \, G H
\right.
\nn\\[2mm]
&& \left. \hspace{5mm} + \, Y^2 + \tfrac{4}{3} \, Y Z + \tfrac{2}{3}\, Z^2 
+ \tfrac{2}{3}\, X^2 + 2\, \eta\, X Z \, \rb I
\nn\\[2mm]
&& \hspace{5mm} = \lb C_I + \tfrac{2}{3} \, C_B \rb I = I \;.
\label{unit.10}
\eea

\begin{figure}[!h]
\begin{center}
\includegraphics[width=1\columnwidth,angle=0]{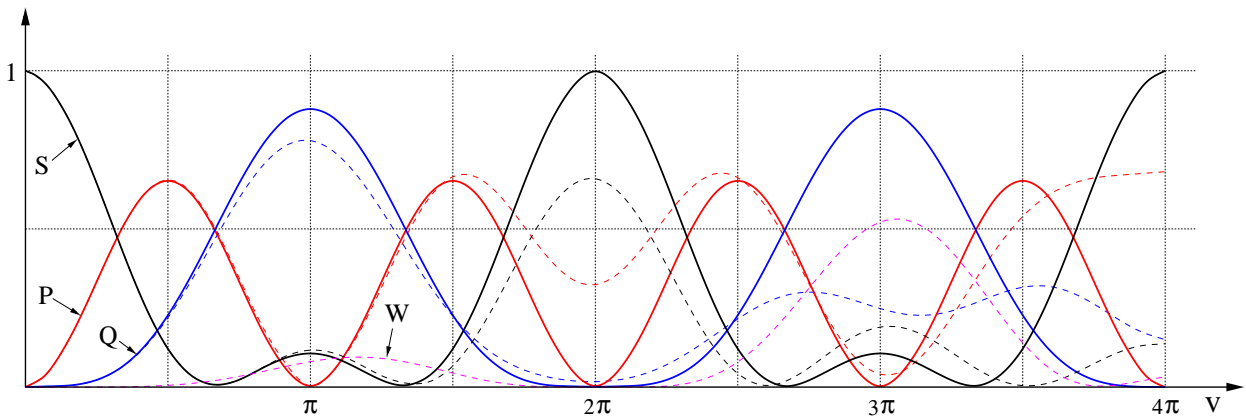}
\caption{$S^2$, $Q^2$, $W^2$ and $P^2$ as functions of $v$ for $\eta=0$ 
(continuous curves) and $\eta=0.1361$ (dashed curves).}
\label{U5}
\end{center}
\end{figure}

This means that 
\bea
U^\dagger \, U = I \rar    S^2 +  Q^2 + W^2 +  P^2  = 1 \;,
\label{unit.8}
\eea
indicating that  the variables $S, Q, W, P$ 
are constrained to the surface 
of a four-dimensional sphere, irrespective of the values of the free 
parameters $v$  and $\eta$.
This is relevant for applications of chiral  symmetry to 
low-energy strong systems, which involve both vector and axial transformations.
In $SU(3)$, the former promote changes in the labels of $Q_i$ and $P_i$ while keeping
$Q= \pm \sqrt{Q_i\,Q_i}$ and $P= \pm \sqrt{P_i\,P_i}$ invariant.
The latter, on the other hand, modify all functions $S, Q, W, P$ together
and the constraint imposed by unitarity corresponds to a generalization of the 
$SU(2)$ condition $\s^2 + \p^2 = \text{constant}$ of the 
non-linear $\s$-model\cite{GellMannLevy}.
The dependence of the functions $S^2$, $Q^2$, $W^2$, and $P^2$ on $v$ 
is displayed in fig.\ref{U5}, where full and dashed curves correspond to 
$\eta=0$  and the arbitrary value $\eta = 1/\sqrt{54}=0.1361$, respectively. 
As expected from the explicit results for $F, \cdots Z$ in eqs.(\ref{anf.46})-(\ref{anf.51}),
just the case $\eta=0$ yields cyclic structures.
It is worth noting that, in this case, the odd scalar 
term $W $ vanishes.

\begin{figure}[!h]
\begin{center}
\includegraphics[width=1\columnwidth,angle=0]{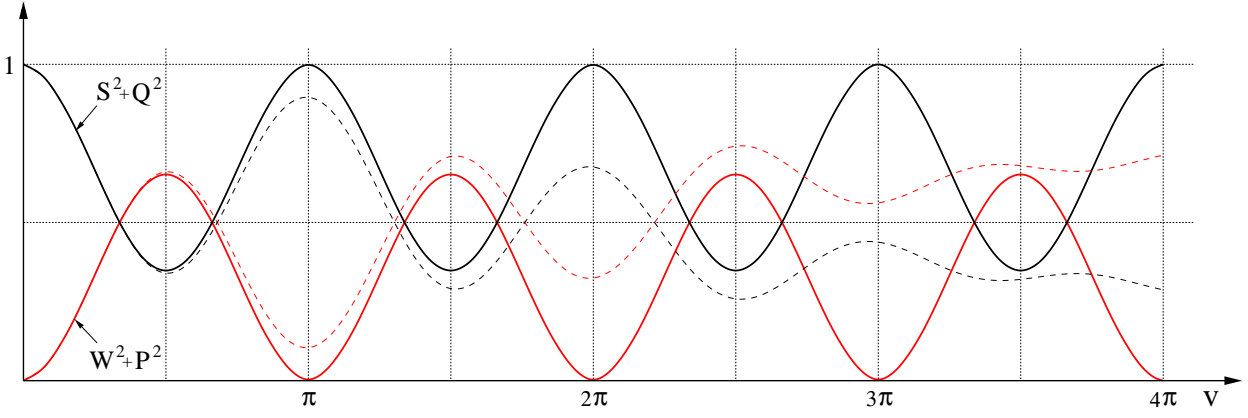}
\caption{$\lp S^2+Q^2\rp$ and $\lp W^2+P^2 \rp $ as functions of $v$ for $\eta=0$ 
(continuous curves) and $\eta=0.1361$ (dashed curves);
note that $(S^2+Q^2)+(W^2+P^2)=1$, as in eq.(\ref{unit.8}). }  
\label{U6}
\end{center}
\end{figure}

The situation in $SU(3)$ contrasts with  the $SU(2)$ case, 
where the variation of the chiral angle $\theta$ gives rise to oscillations
of scalar and pseudoscalar variables  constrained to a circle.
Denoting the trace by $\la \cdots \ra$, one shows in fig.\ref{U6} the behaviour 
of the components
% $U_\text{even}^2=S^2 +Q^2 $ and $U_\text{odd}^2=P^2$ 
$S^2 +Q^2 =\tfrac{1}{3} \la U_\text{even}^2 \ra $ and 
$(W^2+P^2) = \tfrac{1}{3} \la U_\text{odd}^2 \ra$ 
as functions of $v$, for $\eta=0$  and $\eta =0.1361$.
In the case $\eta=0$ one has $W=0 $ and these functions oscillate, 
with values restricted to the intervals 
$1 \geq \lp S^2+Q^2\rp \geq 1/3$ and $2/3 \geq P^2 \geq 0$.
% $1 \geq U_\text{even}^2 \geq 1/3$ and $2/3 \geq U_\text{odd}^2 \geq 0$.
The idividual scalar contributions $S^2$ and $Q^2$ do vanish at specific points,
but their sum does not.
This interplay between $S$ and $Q$ within the even sector 
is a distinctive feature of the $SU(3)$ case.

%--------------------------------------------------------------------------------
\section{left and right forms}
\label{s5}

The analytic result for $U$, eq.(\ref{anf.57}), allows one to derive the left and right 
forms $L^\m$ and $R^\m$, defined by
\bea
L^\m = U^\dagger \, \frac{\dr U}{\dr x^\m} \;,
\hspace{10mm} 
R^\m = U \, \frac{\dr U^\dagger}{\dr x^\m} \;.
\label{left.1}
\eea
They  are related to the vector and axial currents $V^\m$ and $A^\m$ by
\bea
&& L^\m = i\, \lp V^\m - A^\m \rp \;,
\hspace{10mm}
R^\m = i\, \lp V^\m + A^\m \rp 
\label{left.2}
\eea
and, owing to the unitarity condition $U^\dagger\,U=I$, one has
\bea
[L^\m]^\dagger = - L^\m\,,
\hspace{10mm}  
[R^\m]^\dagger = - R^\m\;.
\label{left.3}
\eea
The left form is evaluated in App.\ref{left} and reads
\bea
&&  L^\m =  i \, \lc \tfrac{3}{2} \lb  \lp Q_i\, \dr^\m Q_j 
+ P_i\, \dr^\m  P_j  \rp  f_{ijk} \rb 
\right.
\nn\\[2mm]
&& \hspace{5mm} \left.
+\,   \lb \sqrt{\tfrac{3}{2}} \lp \,S\, \dr^\m P_k  - P_k\, \dr^\m S 
- W\, \dr^\m Q_k + Q_k \, \dr^\m W   \rp 
\right.\right.
\nn\\[2mm]
&& \hspace{5mm} \left. \left.
+ \tfrac{3}{2} \lp  Q_i\, \dr^\m P_j 
- P_i\, \dr^\m Q_j \,\rp d_{ijk} \rb\rc \l_k \;.
\label{left.4}
\eea
Writing $L^\m = i\, \lp V_k^\m - A_k^\m \rp \l_k $, eqs.(\ref{D.38}) and (\ref{D.40})
allow one to express the currents in terms of the basic functions $F, \cdots ,Z$ as
\bea 
&&  V_k^\m =  \lb \lp H^2 + X^2 \rp \, \hat{v}_i \, \dr^\m \hat{v}_j 
+ \, \tfrac{1}{\sqrt{3}}  \lp G\, H + X\, Z \rp 
\lp  \hat{v}_i \, \dr^\m \hat{b}_j + \hat{b}_i \, \dr^\m \hat{v}_j \rp
\right. 
\nn\\[2mm]
&& \hspace{5mm} \left.
+ \, \tfrac{1}{3} \lp G^2 + Z^2 \rp \, \hat{b}_i \, \dr^\m \hat{b}_j  \rb  f_{ijk}  \;,
\label{left.5}\\[4mm]
&& A_k^\m = - \lc \lb 1 \rb  \hat{v}_k \,\dr^\m v 
\right.
\nn\\[2mm]
&& \left.
+ \, \frac{1}{(1-\tfrac{27}{4}\eta^2)} 
\lb  \lp G Y - F Z \rp  + \tfrac{9}{4} \eta \lp F X - H Y \rp 
\right.\right.
\nn\\[2mm]
&& \left.\left.
+ \,\tfrac{3}{4} \eta \lp HZ-G X \rp - \tfrac{9}{4}\, v\, \eta  
\rb  \hat{v_k}  \, \dr^\m  \eta 
\right.
\nn\\[2mm]
&& \left.
+ \, \frac{1}{ (1-\tfrac{27}{4}\eta^2)} \, \tfrac{1}{\sqrt{3}}
\lb  \tfrac{3}{2} \lp H Y - F X \rp  + \tfrac{1}{2} \lp G X - H Z \rp
\right.\right.
\nn\\[2mm]
&& \left.\left. 
+ \, \tfrac{9}{2} \eta \lp F Z - G Y \rp 
+  \tfrac{3}{2} \,  v
\rb   \hat{b}_k  \, \dr^\m \eta 
\right.
\nn\\[2mm]
&& \left. 
+ \lb \lp F + \tfrac{2}{3} G \rp  X - H \lp Y + \tfrac{2}{3}  Z \rp \rb \dr^\m \hat{v}_k 
\right.
\nn\\[2mm]
&& \left. 
+ \,  \tfrac{1}{\sqrt{3}}  \lb \lp F +\tfrac{2}{3}  G \rp  Z - G \lp Y + \tfrac{2}{3} Z \rp \rb 
\dr^\m \hat{b}_k
\right.
\nn\\[2mm]
&& \left. 
+ \tfrac{1}{\sqrt{3}}  \lp H Z -  G X \rp 
\lp \hat{v}_i \, \dr^\m \hat{b}_j - \hat{b}_i \, \dr^\m \hat{v}_j \rp  d_{ijk} \rc \;.
\label{left.6}
\eea

%-------------------------------------------------------------------------------
\section{classical limit}
\label{s4}

\begin{figure}[!h]
\begin{center}
\includegraphics[width=0.5\columnwidth,angle=0]{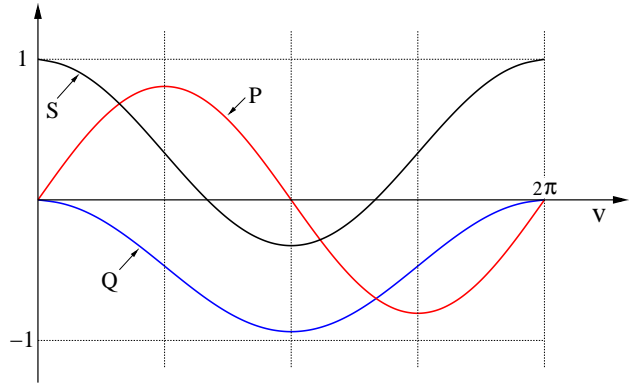}
\caption{Classical $S$, $Q$ and $P$ as functions of $v$.}
\label{U1}
\end{center}
\end{figure}

In the case of spontaneous symmetry breaking,
the variable $v$ may acquire a non-vanishing vacuum expectation value and
become the $SU(3)$ analogous of the $SU(2)$ chiral angle $\theta$.
As the same does not apply for $\eta$, which has odd parity under $\bv \rar -\bv$,
one referers to the situation $\la 0| v |0\ra \neq 0$ and $\eta \rar 0$
as the classical limit.
In this case,  one has
\bea
&& k_1 \rar \eta 
\hspace{5mm} \rar \hspace{5mm} 
\a_1 \rar -\eta^2 \;,
\label{class.1}\\[1mm]
&& k_{2,3} \rar \pm 1 +  \frac{\eta}{2} 
\hspace{5mm} \rar \hspace{5mm} 
\a_{2,3} \rar   - (1 \pm \eta) \;,
\label{class.2}
\eea
and finds
\bea
&& F \rar 1 \;,
\label{class.3}\\[1mm]
&&G \rar -1 + \cos v \;,
\label{class.4}\\[1mm]
&& X \rar \sin v \;,
\label{class.5}
\eea
whereas $ H, Y, Z \rar \cO(\eta)$.

 \begin{figure}[!b]
\begin{center}
\includegraphics[width=1\columnwidth,angle=0]{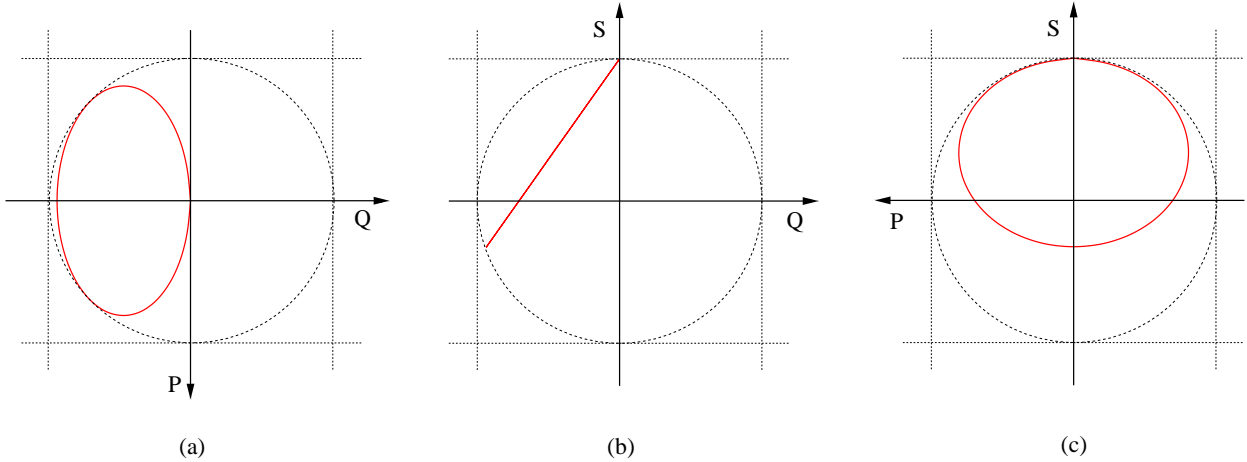}
\caption{Projections of the classical circle (in red) 
over planes  (a) $PQ$, (b) $QS $, and (c) $ SP $; 
figure (b) shows the profile of the circle over the plane 
associated with $U_\text{even} $, whereas figures (a) and (c) 
are obtained by rotating it by $\pi/2$ along axes $Q $ and $S $ respectively;
in all figures, the axis not shown points out of the page.}
\label{U2}
\end{center}
\end{figure}

\begin{figure}[!h]
\begin{center}
\includegraphics[width=0.7\columnwidth,angle=0]{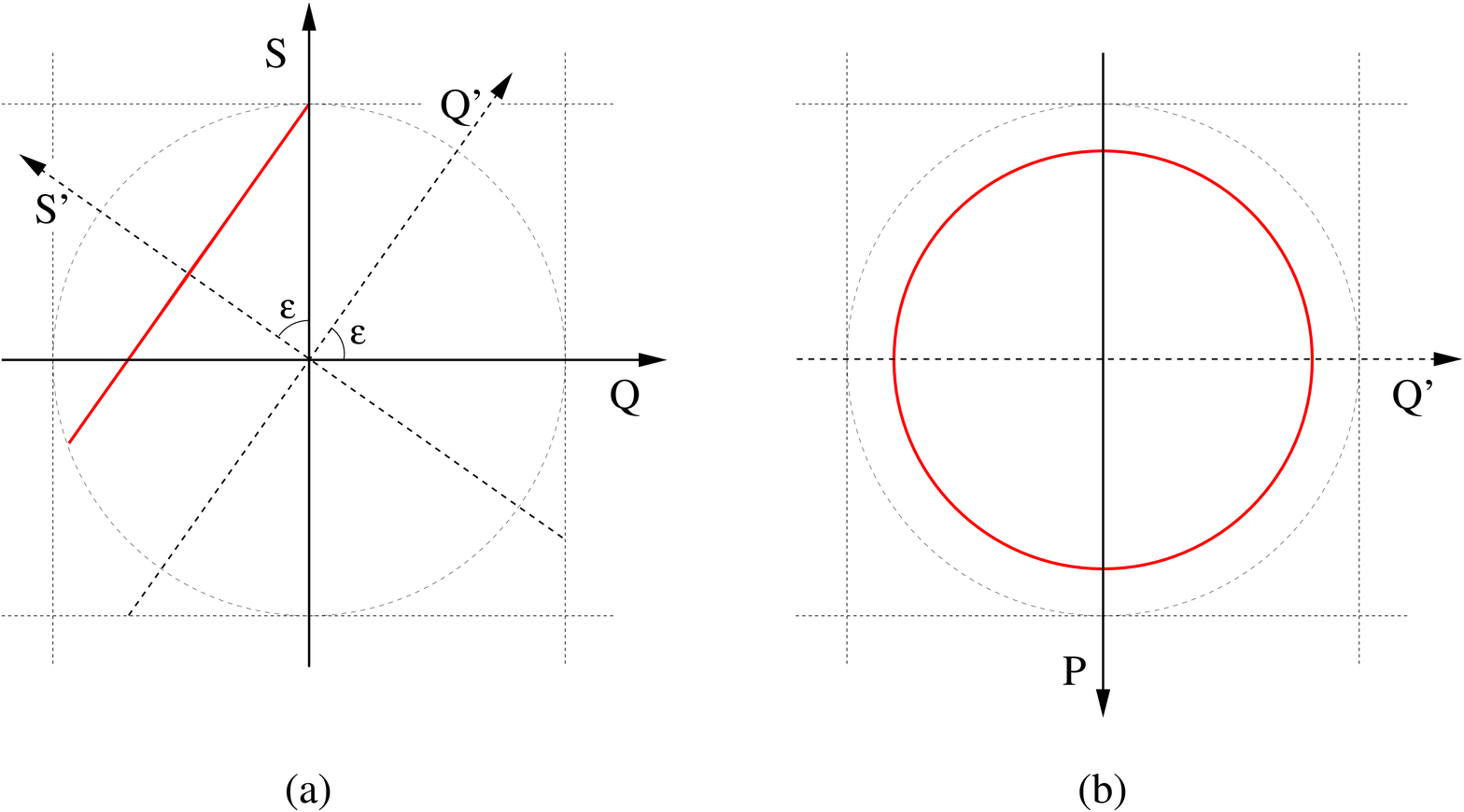}
\caption{Projections of the classical circle (in red) over planes (a)  $Q'S'$ and (b) $Q'P$;
in both figures, the axis not shown points out of the page. }
\label{U3}
\end{center}
\end{figure}

In the classical limit $W\rar 0$ and the behavior of the functions
\bea
&& S \rar \tfrac{1}{3} \, \lp 1 + 2\, \cos v \rp \;,
\label{class.6}\\[1mm]
&& Q \rar \tfrac{\sqrt{2}}{3} \, \lp -1 + \cos v \rp \;,
\label{class.7}\\[1mm]
&& P \rar  \sqrt{\tfrac{2}{3}} \, \sin v \;,
\label{class.8}
\eea
is shown in fig.\ref{U1} and 
the matrix $U$ becomes
\bea 
U = \lb \tfrac{1}{3} \lp 1 + 2\, \cos v \rp \, I
+ \tfrac{1}{\sqrt{3}}\, \lp -1 + \cos v  \rp \, \hat{\bb}\cdot \bl \rb 
+ i\, \lp \sin v \rp \hat{\bv}\cdot \bl \;.
\label{class.9}
\eea
The unitarity condition (\ref{unit.8}) constrains 
$S$, $Q$ and $P$ to the surface of a sphere, since
\bea
U^\dagger \,U  = I   \rar   S^2 + Q^2 + P^2 = 1 \;,
\label{class.10}
\eea
and the variation of $v$ gives rise to a circumference,   
with projections over planes $PQ$,  $Q S $, and $ SP $
shown in fig.\ref{U2}. 
Figure $(b)$, depicting the two components of $U_\text{even}$,
is particularly interesting, for it shows the profile of a circle  as
a straight line, for eqs.(\ref{class.6}) and (\ref{class.7}) yield
\bea
S = 1 - \sqrt{2}\, Q \;.
\label{class.11}
\eea
Thus, the path determined by $v $ is tilted circumference, defined by the intersection of 
the unit sphere with a plane orthogonal to the axes $Q $ and $S $, 
inclined by an angle $\e = \tan^{-1} \sqrt{2}$,
which amounts to $\sin\e=\sqrt{2/3}$, $\cos \e=  \sqrt{1/3}$, 
and $\e\sim 54.76^\circ$. 
Performing a rotation around the $P$ axis,   as in fig.\ref{U3},    one has
\bea
&&  S' = \sqrt{\tfrac{1}{3}} \, S - \sqrt{\tfrac{2}{3}} \, Q \;,
\hspace{15mm}
Q' = \sqrt{\tfrac{2}{3}} \, S + \sqrt{\tfrac{1}{3}} \, Q \;,
\label{class.12}
\eea
and the equation of the plane containing the circle is $S' = \sqrt{1/3}$.
Its edge is determined by condition  (\ref{class.10}),
which now reads $Q^{'2} + P^2 = 1 - S^{'2}=2/3$, 
corresponding to a radius of $\sqrt{2/3}$ 
and to $Q' = \sqrt{2/3}\, \cos v $.

%-------------------------------------------------------------------------------
\section{chiral symmetry}
\label{s6}

One now concentrates on the case of pseudoscalar mesons $\bphi$ 
and, making  $\bv \rar \bphi$,
discusses  the chiral transformations of the matrix $U(\bfi)$ given by  eq.(\ref{anf.57}).
Its vector transformations are associated with changes in the directions of  $ \hat{\bphi}$
and $ \hat{\bb}$ and need not be written out explicitly.   
Concerning axial tranformations $\d^A \bphi$,
the most general non-linear form has been discussed by Weinberg\cite{Weinberg68}
and is given by
\bea
\d^A \phi_a = f^A(\phi^2) \, \b_a + g^A(\phi^2) \, \b_i \,\phi_i \,\phi_a \;,
\label{chis.1}
\eea
where $\b_i$ are free parameters, $f^A$ is an arbitrary function and
\bea
&& g^A = \frac{2 f^A f^{'A}+1}{f^A-2\,\phi^2 f^{'A}} \;,
\label{chis.2}
\eea
with $f^{'A} = df^A/d \phi^2$.
The axial transformation of a generic function $\psi(v,\eta)$ is 
\bea
\d^A \psi = \frac{d \psi}{d \phi_a} \, \d^A \phi_a
= \lb \frac{\dr \psi}{\dr \phi} \, \hat{\phi}_a 
+ \frac{\dr \psi}{\dr \eta} \, \frac{1}{\phi} \lp - 3\eta\, \hat{\phi_a} 
+ \frac{2}{\sqrt{3}} \, \hat{b}_a  \rp \rb \d^A \phi_a \;,
\label{chis.3}
\eea
using eq.(\ref{D.30}).
Evaluating the derivatives of $(F, \cdots Z)$
with the help of  eqs.(\ref{anf.23}), (\ref{anf.24}) and 
(\ref{D.11})-(\ref{D.16}), one has 
\bea
&& \d^A F =\frac{1}{1-\tfrac{27}{4} \eta^2}
\lc \lb \frac{1}{\phi} \lp 9 \, \eta^2 G - 3\, \eta H \rp 
+ \lp \tfrac{9}{2} \, \eta^2  X - 3\,\eta Y - \eta Z \rp \rb \hat{\phi}_a
\right.
\nn\\[2mm]
&& \hspace{8mm} \left.
+ \frac{2}{\sqrt{3}} \lb \frac{1}{\phi} \lp - 3\, \eta G + H \rp 
+ \lp - \tfrac{3}{2} \, \eta X + Y + \tfrac{9}{4}\, \eta^2 Z \rp  \rb \hat{b}_a 
\rc \d^A \phi_a \;,
\label{chis.4} \\[2mm]
%
%G
&& \d^A G =\frac{1}{1-\tfrac{27}{4} \eta^2}
\lc \lb \frac{1}{\phi} \lp - \tfrac{27}{2}\,\eta^2 G + \tfrac{9}{2}\, \eta H \rp 
+ \lp -X + \tfrac{9}{2}\, \eta Y + \tfrac{3}{2}\, \eta Z \rp \rb \hat{\phi}_a
\right.
\nn\\[2mm]
&& \hspace{8mm} \left.
+ \frac{2}{\sqrt{3}} \lb \frac{1}{\phi} \lp \tfrac{9}{2}\, \eta G - \tfrac{3}{2}\, H \rp 
+ \lp \tfrac{9}{4} \, \eta X - \tfrac{3}{2} \, Y - \tfrac{1}{2}\, Z \rp  \rb \hat{b}_a 
\rc \d^A \phi_a \;,
\label{chis.5}\\[2mm]
%
%
%H
&& \d^A H =\frac{1}{1-\tfrac{27}{4} \eta^2}
\lc \lb \frac{1}{\phi} \lp 3\, \eta G - \tfrac{27}{4}\, \eta^2 H \rp 
+ \lp \tfrac{3}{2} \, \eta X - Y - Z + \tfrac{9}{2} \, \eta^2 Z \rp
 \rb \hat{\phi}_a
\right.
\nn\\[2mm]
&& \hspace{8mm} \left.
+ \frac{2}{\sqrt{3}} \lb  \frac{1}{\phi} \lp - G + \tfrac{9}{4}\, \eta H \rp 
+ \lp  -\tfrac{1}{2} \, X + \tfrac{9}{4} \, \eta Y + \tfrac{3}{4} \, \eta Z \rp \rb \hat{b}_a 
\rc \d^A \phi_a \;,
\label{chis.6} 
\eea

\bea
%X
&& \d^A X =\frac{1}{1-\tfrac{27}{4} \eta^2}
\lc \lb \frac{1}{\phi} \lp -\tfrac{27}{4}\, \eta^2 X + 3\, \eta Z \rp 
+ \lp F + G - \tfrac{9}{2} \, \eta^2 G - \tfrac{3}{2} \, \eta H \rp \rb \hat{\phi}_a
\right.
\nn\\[2mm]
&& \hspace{8mm} \left.
+ \frac{2}{\sqrt{3}} \lb  \frac{1}{\phi} \lp \tfrac{9}{4}\, \eta X - Z \rp 
+ \lp -\tfrac{9}{4} \, \eta F - \tfrac{3}{4}\, \eta G + \tfrac{1}{2} \, H \rp  \rb \hat{b}_a 
\rc \d^A \phi_a \;,
\label{chis.7} \\[2mm]
%
%Y
&& \d^A Y =\frac{1}{1-\tfrac{27}{4} \eta^2}
\lc \lb \frac{1}{\phi} \lp -3\,\eta X + 9\, \eta^2 Z \rp 
+ \lp 3\, \eta F + \eta G - \tfrac{9}{2}\, \eta^2 H \rp \rb \hat{\phi}_a
\right.
\nn\\[2mm]
&& \hspace{8mm} \left.
+ \frac{2}{\sqrt{3}} \lb \frac{1}{\phi} \lp X - 3\, \eta Z \rp 
+ \lp -F - \tfrac{9}{4}\, \eta^2 G  + \tfrac{3}{2}\, \eta H\rp  \rb \hat{b}_a
\rc \d^A \phi_a \;,
\label{cs.8} \\[2mm]
%
%Z
&& \d^A Z =\frac{1}{1-\tfrac{27}{4} \eta^2}
\lc \lb \frac{1}{\phi} \lp \tfrac{9}{2}\, \eta X - \tfrac{27}{2} \, \eta^2 Z \rp 
+ \lp - \tfrac{9}{2}\, \eta F - \tfrac{3}{2} \, \eta G + H \rp \rb \hat{\phi}_a
\right.
\nn\\[2mm]
&& \hspace{8mm} \left.
+ \frac{2}{\sqrt{3}} \lb \frac{1}{\phi} \lp - \tfrac{3}{2}\, X + \tfrac{9}{2} \, \eta Z \rp 
+ \lp \tfrac{3}{2}\, F +  \tfrac{1}{2} \, G - \tfrac{9}{4}\, \eta H \rp  \rb \hat{b}_a 
\rc \d^A \phi_a \;,
\label{chis.9} 
\eea
whereas the two directions transform as
\bea
&& \d^A \, \hat{\phi}_i = \frac{1}{\phi} \lp \d_{ia} 
- \hat{\phi}_i \,\hat{\phi}_a \rp \, \d^A \phi_a \;,
\label{chis.10}\\[2mm]
&& \d^A \, \hat{b}_i = 
\frac{2\sqrt{3}}{\phi} \lp  d_{ija} \, \hat{\phi}_j  
- \tfrac{1}{\sqrt{3}}\, \hat{b}_i \,  \hat{\phi}_a \rp \,  \d^A \phi_a \;.
\label{chis.11}
\eea
One notes that, as expected, axial transformations change the parities of 
the functions $F, \cdots, Z$, and of the directions $\hat{\bphi} $ and $\hat{\bb} $,
under the operation $\bphi \rar - \bphi$.

Using results (\ref{chis.4})-(\ref{chis.9}) one can, for instance, 
show that the functions $C_I$, $C_B$  and
$C_A$ given by eqs.(\ref{unit.2})-(\ref{unit.3}) are invariant under 
axial transformations by means of explicit calculations.
In the case of classical fields, these transformations become much simpler and read
\bea
&& \d^A F =  0 
\hspace{5mm} \rar \hspace{5mm} 
\d^A 1 =  0 \;,
\label{chis.12} \\[2mm]
&& \d^A G =  -X \, \hat{\phi}_a \, \d^A \phi_a 
\hspace{5mm} \rar \hspace{5mm} 
\d^A \lp - 1 + \cos \phi\rp = - \sin \phi \, \d^A \phi \;,
\label{chis.13} \\[2mm]
&& \d^A X = \lp F + G \rp \hat{\phi}_a \, \d^A \phi_a 
\hspace{5mm} \rar \hspace{5mm} 
\d^A \sin \phi = \cos \phi \, \d^A \phi  \;
\label{chis.14} 
\eea
using $\hat{\phi}_a \, \d^A \phi_a = \d^A \phi$.
Thus, the axial transformation implements a rotation along the tilted circumference
discussed in sect.\ref{s4}.

%-------------------------------------------------------------------------------
\section{summary}
\label{summa}

One  presents an analytic expression for the $SU(3)$ unitary matrix which,
although motivated by low-energy hadron physiscs, has a more general validity.
\\

\ni
{\bf 1.}
The $SU(2)$ unitary matrix $U$ is well known to have
two equivalent representations, given by
$ U = \exp\lb \, i\, \btau \cdot \hat{\bpi}  \, \theta\,\rb 
= \cos\th + i\, \btau \cdot \hat{\bpi}  \, \sin\th $, 
where $\btau$ are Pauli matrices and   $\bpi = (\p_1, \p_2,\p_3) $
is the pion field. 
In sect.\ref{s2} one extends this result to the $SU(3)$ case and, 
for Gell-Mann matrices $\bl$,  
derives the identity 
\bea 
U = \exp\lb\,  i\, \bv\cdot \bl \, \rb 
= \lb \, S \,I  + \sqrt{\tfrac{3}{2}} \, \bQ \cdot \bl \, \rb
+ i\, \lb \, W \,I  + \sqrt{\tfrac{3}{2}} \, \bP \cdot \bl \, \rb \;,
\nn
\eea
with $ ( S + i W ) = \lb (F + i Y)  + \tfrac{2}{3}\,(G+i Z) \rb $, 
$ ( Q +iP)_i = \sqrt{\tfrac{2}{3}}\, 
\lb (H+ iX) \, \hat{v}_i +\tfrac{1}{ \sqrt{3}}\,(G+iZ)\, \hat{b}_i \rb  $,
$v_i=[\,v_1, \cdots  v_8\,]$, $ b_i = d_{ijk} \, v_j \, v_k $, and
functions $F, \cdots, Z$ given by eqs.(\ref{anf.46})-(\ref{anf.51}),
depending on $v=|\bv|$ and 
$\eta = 2\, d_{ijk} \, \hat{v}_i \, \hat{v}_j \, \hat{v}_k /3 $.
\\

\ni
{\bf 2.} Unitarity constrains the functions 
$S, Q=\pm \sqrt{Q_i Q_i} , \, W, P = \pm \sqrt{P_i P_i}$ to the surface of a 
four-sphere, since
\bea
U^\dagger \,U = I \rar   S^2 +  Q^2 + W^2 +  P^2  = 1 \;,
\nn
\eea
for all values of $v$ and $\eta$.
\\

\ni
{\bf 3.} The analytic result for $U$ allows the explicit evaluation of the left form, 
which reads
\bea
&&  L^\m =  i \, \lc \tfrac{3}{2} \lb  \lp Q_i\, \dr^\m Q_j 
+ P_i\, \dr^\m  P_j  \rp  f_{ijk} \rb 
\right.
\nn\\[2mm]
&& \hspace{5mm} \left.
+\,   \lb \sqrt{\tfrac{3}{2}} \lp \,S\, \dr^\m P_k  - P_k\, \dr^\m S 
-W\, \dr^\m Q_k + Q_k \, \dr^\m W   \rp 
\right.\right.
\nn\\[2mm]
&& \hspace{5mm} \left. \left.
+ \tfrac{3}{2} \lp  Q_i\, \dr^\m P_j 
- P_i\, \dr^\m Q_j \,\rp d_{ijk} \rb\rc \l_k \;.
\nn
\eea
This gives rise to  the right form as well as to vector and axial currents.
In sect.\ref{s5},  one presents expressions in terms of the functions $F, \cdots Z$,
which can be used in calculations. 
\\

\ni
{\bf 4.} In the classical limit, corresponding to $\eta \rar 0$ and 
$\la 0| v |0 \ra \neq 0$, one has $W\rar 0 $ and obtains the simpler form
\bea
&& U =\lb \, S \,I  + \sqrt{\tfrac{3}{2}} \, Q \, \hat{\bb} \cdot \bl \, \rb
+ i\, \sqrt{\tfrac{3}{2}} \, P \, \hat{\bv} \cdot \bl  \;,
\eea
with $ S \rar \tfrac{1}{3} \, \lp 1 + 2\, \cos v \rp $,
$ Q \rar \tfrac{\sqrt{2}}{3} \, \lp -1 + \cos v \rp $ and
$P \rar  \sqrt{\tfrac{2}{3}} \, \sin v $,
satisfying $  S^2 +  Q^2 +  P^2  = 1 $.
The matrix $U $ becomes a cyclic function of $v$
and oscillates, but its even and odd components
under $\bv \rar -\bv$  remain restricted to the intervals 
$1 \geq (S^2 + Q^2) \geq 1/3$ and $2/3 \geq P^2 \geq 0$,
as indicated in fig.\ref{U6}.
The variation of $v$ determines a tilted circumference
with radius $\sqrt{2/3}$ in the space defined by 
$I$, $\hat{\bb}\cdot \bl $, and $\hat{\bv}\cdot \bl $,
illustrated  in fig.\ref{U2}. 
In terms of the variable $ Q' =  2 \, S/\sqrt{3} +  Q /\sqrt{3} $, its 
edge is given by  $Q^{'2} + P^2 =2/3$ and,
in the case of chiral symmetry, this corresponds to a generalization of the 
condition $\s^2 + \pi^2 = \text{constant} $ of the non-linear $SU(2)$ $\s$-model. 
\\

\ni
{\bf 5.} In sect.\ref{s6}, the generic analytic expression for $U$ 
is adapted to low-energy flavor $SU(3)$
by associating the $v_i$  with pseudoscalar fields $\phi_i$ and
one displays its axial transformation properties,
involving both the functions $F, \cdots Z$ and the directions $\hat{\bphi}$ 
and $\hat{\bb}$.
In the classical limit, one has 
$ \d^A \cos \phi = - \sin \phi \, \d^A \phi$ and 
$ \d^A \sin \phi = \cos \phi \, \d^A \phi  $,
indicating that
the axial transformation corresponds to a rotation along the tilted circumference.
\\

\ni
{\bf 6.}  Results given in sects.\ref{s2}, \ref{s3} and \ref{s5} 
are generic and not committed
to a particular interpretation of the variable $\bv$.
Hence,  they may prove to be useful in problems involving 
three degrees of freedom or three state systems.
In QCD, one has the lightest flavors $u, d, s$
and  the basic colors, where it might be instrumental either to study color superconductivity
or to investigate topological properties  of the classical solution, as in the Skyrme model.
The interest of the analytic form of $U$ is not restricted to hadron
physics and it may also be applied in other areas, such as optics, 
geometric phases, quantum computation, and communication.
\\

%---------------------------------------------------------------------------------
\appendix 
\section{auxiliary functions}
\label{auxfunc}

The explicit components of the vector $\bb$ are given by
\bea
&& b_1 = \tfrac{2}{\sqrt{3}} \, v_1\, v_8 + v_4\, v_6 + v_5\, v_7 \;,
\label{A.1}\\[2mm]
&& b_2= \tfrac{2}{\sqrt{3}} \, v_2 \, v_8 - v_4 \, v_7 + v_5 \, v_6 \;,
\label{A.2}\\[2mm]
&& b_3 = \tfrac{2}{\sqrt{3}} v_3\, v_8 
+ \tfrac{1}{2}\,\lp  v_4^2 + v_5^2 - v_6^2 - v_7^2 \rp \;,
\label{A.3}\\[2mm]
&& b_4 = v_1\, v_6  - v_2\, v_7 + v_3 \, v_4 -\tfrac{1}{\sqrt{3}}\, v_4\, v_8 \;,
\label{A.4}\\[2mm]
&& b_5 = v_1\, v_7  + v_2\, v_6 + v_3 \, v_5 -\tfrac{1}{\sqrt{3}}\, v_5\, v_8 \;,
\label{A.5}\\[2mm]
&& b_6 = v_1\, v_4  + v_2\, v_5 - v_3 \, v_6 -\tfrac{1}{\sqrt{3}}\, v_6\, v_8 \;,
\label{A.6}\\[2mm]
&& b_7 = v_1\, v_5  - v_2\, v_4 - v_3 \, v_7 -\tfrac{1}{\sqrt{3}}\, v_7\, v_8 \;,
\label{A.7}\\[2mm]
&& b_8 = \tfrac{1}{\sqrt{3}} \lb \, v_1^2 +  v_2^2 +  v_3^2 
- \tfrac{1}{2}  \lp v_4^2 + v_5^2 + v_6^2 + v_7^2 \rp 
- v_8^2 \rb \;,
\label{A.8}
\eea
whereas the function $D$ reads
\bea
&& \hspace{-5mm} D = d_{ijk} \, v_i\,v_j\,v_k
\nn\\[2mm]
&&  = \sqrt{3} \,  \lb \, v_1^2 + v_2^2 + v_3^2 
- \tfrac{1}{2} \lp v_4^2 + v_5^2 + v_6^2 + v_7^2 \rp 
- \tfrac{1}{3}\, v_8^2 \, \rb \, v_8
\nn\\[2mm]
&& + \;  3\,v_1 \lp \,v_4 \,v_6 + v_5\,v_7\,\rp  
+ 3\,v_2 \lp \,- v_4 \,v_7 + v_5\,v_6 \,\rp  
+\tfrac{3}{2}\,  v_3  \lp \, v_4^2 + v_5^2 -  v_6^2 - v_7^2 \rp  \;.
\label{A.9}
\eea

Using Jacobi identities\cite{Gasio}, one shows that the
components of $\bb$ satisfy the conditions
\bea
&& f_{ijk} \, v_j \, b_k = 0\;,
\label{A.10}\\[2mm]
&& d_{jks} \, v_j \, b_k = \tfrac{1}{3}\, v^2 \, v_i \;.
 \label{A.11}
\eea
Alternatively, it is straightforward to prove these results 
by using directly eqs.(\ref{A.1})-(\ref{A.8}).
Multiplying eq.(\ref{A.11}) by $b_i$, one finds $b^2 = v^4/3 $.
Also, using $B\,B = B + \eta\, A$, eq.(\ref{anf.13}), one has 
$(\hat{b}\cdot\l) \, (\hat{b} \cdot \l) = 2/3 + 3\,\eta\, \hat{v}\cdot \l - \hat{b}\cdot \l/ \sqrt{3}$, 
which yields
\bea
&& d_{ijk} \, b_j \, b_k = \eta\, v^3 \, v_i - \tfrac{1}{3}\, v^2 \, b_i \;.
 \label{A.12}
\eea

%---------------------------------------------------------------------------------
\section{differential equation}
\label{cubic}

One considers the differential equation (\ref{anf.27}), that reads
\bea
&& \frac{\dr^6F}{\dr v^6}  + 2\, \frac{\dr^4F}{\dr v^4}  + \frac{\dr^2F}{\dr v^2}   
+ \eta^2 \,F  =0 \;.
\label{B.1}
\eea
Its solution has the general form $F = \exp(q\,v) $\;,
where $q$ satisfies the algebraic equation
\bea
q^6 + 2\,q^4 + q^2 + \eta^2 =0 \;.
\label{B.2}
\eea
Defining $\a=q^2$, one has the cubic equation
\bea
\a^3 + 2\, \a^2 + \a + \eta^2 =0\;,
\label{B.3}
\eea
which has the solutions
\bea
&& \a_1 = q_1^2 =  -\tfrac{2}{3} \lb \, 1 - \cos(\th/3) \, \rb \;,
\label{B.4}\\[2mm]
&& \a_2 = q_2^2 =  -\tfrac{2}{3} \lb \, 1 - \cos(\th/3+ 2\pi/3) \, \rb \;,
\label{B.5}\\[2mm]
&& \a_3 = q_3^2 =  -\tfrac{2}{3} \lb \, 1 - \cos(\th/3-2\pi/3) \, \rb \;,
\label{B.6}
\eea 
with 
\bea
&& \cos(\theta) = 1 - 27\,\eta^2/2 \;,
\label{B.7}\\[2mm]
&& \sin(\theta) = 3\sqrt{3}\, \eta \, \sqrt{1-27\,\eta^2/4} \;.
\label{B.8}
\eea
As $\a_i<0$, one defines $q_i = i\, k_i$, and has
\bea
&& k_1 = \tfrac{2}{\sqrt{3}}\, \sin (\th/6) \;,
\label{B.9}\\[2mm]
&& k_2 =  \tfrac{2}{\sqrt{3}}\, \sin (\th/6+\pi/3)
=  \cos(\theta/6) + \tfrac{1}{\sqrt{3}} \,\sin(\th/6) \;,
\label{B.10}\\[2mm]
&& k_3 =  \tfrac{2}{\sqrt{3}}\, \sin (\th/6-\pi/3)
= - \cos(\theta/6) + \tfrac{1}{\sqrt{3}} \,\sin(\th/6) \;.
\label{B.11}
\eea
The function $F$ is real and its most general form reads
\bea
F = \b_1 \, \cos(k_1\,v) + \b_2 \, \cos(k_2\,v) + \b_3 \, \cos(k_3\,v) \;,
\label{B.12}
\eea 
where the $\b_i$ are constants.

The $k_i$ satisfy the constraints
\bea
&& k_1 = k_2 + k_3  \;,
\label{B.13}\\[2mm]
&& k_1^2 + k_2^2 + k_3^2 =2 \;,
\label{B.14}\\[2mm]
&& k_1\, k_2\, k_3 = - \eta \;,
\label{B.15}
\eea
whereas, for the roots of the cubic equation (\ref{B.3})  one has the usual conditions
\bea
&& \a_1 + \a_2 + \a_3 = -2 \;,
\label{B.16}\\[2mm]
&& \a_1\,\a_2 + \a_2\, a_3 + \a_3\,\a_1 = 1 \;,
\label{B.17}\\[2mm]
&& \a_1 \, \a_2 \, \a_3 = -\eta^2 \;,
\label{B.18}
\eea 
Combining (\ref{B.16}) and (\ref{B.17}), one finds the useful result
\bea
&& \a_i^2 + 2\,\a_i + \a_j^2 + 2\, \a_j  + \a_i \, \a_j + 1 = 0 \;,
\label{B.19}
\eea
that can also be rewritten as 
\bea
&& 
(\a_i^2+\a_i) + (\a_j^2 +\a_j) = - (\a_i +1)\,(\a_j +1) \;.
\label{B.20}
\eea
Multiplying it by $\a_i$ and using (\ref{B.3}), one gets
\bea
\a_i^2 \, \a_j + 2\, \a_i\,\a_i + \a_i \, \a_j^2 - \eta^2
= \a_i \, \a_j \,\lp \, 2+ \a_i + \a_j \,\rp - \eta^2 = 0 
\label{B.21}
\eea
and, using (\ref{B.19}) and (\ref{B.21}), one also shows that
\bea
 (\a_i^2 + \a_i) \, (\a_j^2 + \a_j) = \eta^2\,(1 + \a_i + \a_j) \;.
\label{B.22}
\eea

%--------------------------------------------------------------------------------
\section{unitarity - proof}
\label{unitproof}

The unitarity of the matrix $U$ is indicated in eq.(\ref{unit.1})
 and here one proves the validity of conditions (\ref{unit.5}). 
Using the shorthands $c_i = \cos(k_i\,v)$ and $s_i= \sin(k_i\,v)$ 
in eqs.(\ref{anf.46})-(\ref{anf.51}) 
and results from App.\ref{cubic}, one has 
\bea
&&  F^2 = \lc \b_1^2 c_1^2\,[1]  + 2\b_1\b_2  c_1 c_2 \, [1]+\cdots \rc \;,
\label{C.1}\\[2mm]
&& G^2 = \frac{1}{\eta^2} 
\lc \b_1^2 c_1^2\, [-\a_1] 
+ 2\b_1 \b_2  c_1 c_2 \, [1 +\a_1 +\a_2]  +\cdots \rc \;,
\label{C.2}\\[2mm]
&& H^2 = \frac{1}{\eta^2} 
\lc \b_1^2 c_1^2 \, [\a_1^2] 
+ 2\b_1 \b_2 c_1 c_2 \, [\a_1 \,\a_2]  +\cdots \rc \;,
\label{C.3}
\eea
\bea 
&& 2 F G = \frac{1}{\eta^2} 
\lc \b_1^2 c_1^2 \, [ 2\, (\a_1^2 +\a_1) ] 
+ 2 \b_1 \b_2 c_1 c_2 \,[- (\a_1 +1)\,(\a_2+1)] +\cdots \rc \;,
\label{C.4}\\[2mm]
&& 2FH = \frac{1}{\eta} 
\lc \b_1^2 c_1^2 \, [ - 2\, \a_1 ] 
+ 2\b_1 \b_2 c_1 c_2   \, [- (\a_1 +\a_2)] +\cdots \rc \;,
\label{C.5}\\[2mm]
&&  2 G H = \frac{1}{\eta} 
\lc \b_1^2 c_1^2 \lb \frac{2\,\a_1}{(\a_1+1)} \rb  
+ 2 \b_1 \b_2 c_1 c_2 \, [-1]  +\cdots \rc \;,
\label{C.6}
\eea

\bea
&& X^2 = \frac{1}{\eta^2} 
\lc \b_1^2\, s_1^2 \, [ \a_1^2 ]  
+ 2 \b_1 \b_2  k_1 k_2 s_1 s_2 \, [1 +\a_1 +\a_2]  +\cdots \rc \;,
\label{C.7}\\[2mm]
&& Y^2 = 
\lc \b_1^2 s_1^2 \, [ 1] 
+  2 \b_1 \b_2  k_1 k_2 s_1 s_2 \, \lb \frac{(\a_1+1)\,(\a_2+1)}{\eta^2} \rb +\cdots \rc \;,
\label{C.8}\\[2mm]
&& Z^2 = \frac{1}{\eta^2} 
\lc \b_1^2 s_1^2 \, [ -\a_1]  
+  2 \b_1 \b_2  k_1 k_2 s_1 s_2 \, [1 ] +\cdots \rc \;,
\label{C.9}
\eea

\bea
&&  2 X Y = \frac{1}{\eta} 
\lc \b_1^2 s_1^2 \, [-2 \,\a_1 ] 
+  2 \b_1 \b_2  k_1 k_2 s_1 s_2 \, \lb -1 + \frac{(\a_1 +1)\,(\a_2+1)}{\eta^2}\rb +\cdots \rc \;,
\label{C.10}\\[2mm]
&&  2 X Z = \frac{1}{\eta} 
\lc \b_1^2 s_1^2 \, \lb \frac{2\,\a_1}{(\a_1+1)}  \rb  
+  2 \b_1 \b_2  k_1 k_2 s_1 s_2 \, \lb -\, \frac{(\a_1 +1)\,(\a_2+1)}{\eta^2}\rb +\cdots \rc \;,
\label{C.11}\\[2mm]
&& 2 Y Z = \frac{1}{\eta^2} 
\lc \b_1^2 s_1^2 \, [ 2\,(\a_1^2+\a_1) ] 
+  2 \b_1 \b_2  k_1 k_2 s_1 s_2 \lb - (2+\a_1 +\a_2) \rb +\cdots \rc \;,
\label{C.12}
\eea

Explicit calculations together with eqs.(\ref{anf.32})-(\ref{anf.34}) 
and (\ref{anf.44}) yield
\bea
&& C_I = F^2 + 2\eta\,GH + Y^2 + 2\eta\, XZ
\nn\\[2mm]
&& \hspace{5mm} 
= \sum \b_i^2 \lb \frac{ 3\, \a_i +1}{\a_i + 1} \rb = \sum \b_i = 1 \;,
\label{C.13}\\[2mm]
&& C_B = G^2 + H^2 +2FG + X^2 + Z^2 + 2YZ 
\nn\\[2mm]
&& \hspace{5mm}
= \frac{1}{\eta^2} \,  \sum \b_i^2  \lb  3\, \a_i^2 + \a_i \rb
= \frac{1}{\eta^2} \,\sum \b_i \, \lp \a_i^2 + \a_i \rp  = 0 \;,
\label{C.14}\\[2mm]
&& C_A = \eta\, G^2 + 2 F H + 2 G H + \eta\, Z^2 + 2 X Y + 2 X Z 
\nn\\[2mm]
&& \hspace{5mm}
=  \frac{1}{\eta} \,  \sum \b_i^2  \lb - \frac{ 3\, \a_i^2 + \a_i}{\a_i + 1} \rb
= - \frac{1}{\eta} \,\sum \b_i \, \a_i = 0 \;.
\label{C.15}
\eea

%--------------------------------------------------------------------------------
\section{left form}
\label{left}

Direct evaluation of $L^\m$ by means of eqs.(\ref{anf.57}), (\ref{left.1}), 
and  condition (\ref{left.3}), yields
\bea
&&  L^\m = i \, \lb S\, \dr^\m W - W\, \dr^\m S + Q_i \, \dr^\m P_i - P_i \, \dr^\m Q_i \rb
\nn\\[2mm]
&& \hspace{5mm} +\, i \, \lc \tfrac{3}{2} \lb  \lp Q_i\, \dr^\m Q_j 
+ P_i\, \dr^\m  P_j  \rp  f_{ijk} \rb 
\right.
\nn\\[2mm]
&& \hspace{5mm} \left.
+\,   \lb \sqrt{\tfrac{3}{2}} \lp \,S\, \dr^\m P_k  - P_k\, \dr^\m S 
-W\, \dr^\m Q_k + Q_k \, \dr^\m W   \rp 
\right.\right.
\nn\\[2mm]
&& \hspace{5mm} \left. \left.
+ \tfrac{3}{2} \lp  Q_i\, \dr^\m P_j 
- P_i\, \dr^\m Q_j \,\rp d_{ijk} \rb\rc \l_k \;.
\label{D.1}
\eea
In the sequence, one shows that the first term of this expression vanishes
and evaluates the other ones in terms of the functions $F, \cdots , Z$.
This requires a set of auxiliary results, presented below.

\subsection{derivatives with respect to $\eta$}

For any function $\psi(v, \eta) $, one has 
\bea
\frac{\dr \psi}{\dr x^\m} = \frac{\dr \psi}{\dr v_a} \;\frac{\dr v_a}{\dr x^\m} 
= \lp \frac{\dr \psi}{\dr v} \,\frac{\dr v}{\dr v_a} 
+  \frac{\dr \psi}{\dr \eta} \, \frac{\dr \eta}{\dr v_a} \rp \frac{\dr v_a}{\dr x^\m} \;.
\label{D.2}
\eea
Derivatives with respect to $v$ are given by eqs.(\ref{anf.23})-(\ref{anf.24}),
whereas for $\dr \psi/\dr \eta$ one uses
\bea
&&  \frac{d\a_i}{d\eta} = - \, \frac{2\,\eta}{(\a_i+1)\, (3\,\a_i +1)} \;,
\label{D.3}\\[2mm]
&& \frac{d k_i}{d\eta} = - \, k_i\, \frac{\eta}{\a_i\, (\a_i+1)\, (3\,\a_i +1)} \;,
\label{D.4}
\eea
together with $c_i = \cos(k_i \, v)$, $s_i= \sin(k_i \, v)$, and obtains
\bea
&& \frac{\dr F}{\dr \eta}  
= \frac{\eta}{(3\a_1+1)^3} \lb \frac{4}{(\a_1 + 1)} \rb c_1 
+ \frac{ v\, \eta }{(3\a_1 +1)^2 \, \a_1} \; k_1  s_1  + (1 \rar 2, 3) \;,
\label{D.5}\\[2mm]
&& \frac{\dr  G}{\dr \eta}  
= -\, \frac{\eta}{(3\a_1 +1)^3} \lb \frac{6}{(\a_1 + 1)} \rb  c_1 
-  \frac{v\, \eta}{(3\a_1 +1)^2\,\a_1\,(\a_1+1)} \, k_1 s_1 + (1 \rar 2, 3) \;,
\label{D.6}\\[2mm]
&& \frac{\dr H}{\dr \eta}  
= -\, \frac{1}{(3\a_1 +1)^3} \lb (3\, \a_1 - 1) \rb c_1
- \frac{v}{ (3\a_1 +1)^2} \, k_1 s_1 + (1 \rar 2, 3) \;,
\label{D.7}\\[2mm]
&& \frac{\dr X}{\dr \eta}  
=  -\,\frac{\eta}{(3\a_1+1)^3} \lb \frac{(3\, \a_1 - 1)}{\a_1\,(\a_1+1)} \rb  k_1 s_1
- \frac{v\, \eta}{(3\a_1+1)^2 \,(\a_1+1)} \; c_1 + (1 \rar 2, 3) \;,
\label{D.8}\\[2mm]
&& \frac{\dr Y}{\dr \eta}  
=  -\, \frac{1}{(3\a_1+1)^3} \lb 4 \rb  k_1 s_1
- \frac{v\, (\a_1+1)}{(3\a_1+1)^2} \; c_1 + (1 \rar 2, 3) \;,
\label{D.9}\\[2mm]
&& \frac{\dr Z}{\dr \eta}  
=  \frac{1}{(3\a_1+1)^3} \lb  6  \rb k_1 s_1
+ \frac{v}{(3\a+1)^2} \; c_1 + (1 \rar 2, 3) \;.
\label{D.10}
\eea
Employing eqs.(\ref{anf.46})-(\ref{anf.51}), one reexpresses these results as
\bea
&& \frac{\dr F}{\dr \eta}  
= \lb \lp - 3\, \eta G + H \rp 
+ v \lp - \tfrac{3}{2} \eta  X +Y + \tfrac{9}{4} \eta^2\, Z \rp \rb 
/(1-\tfrac{27}{4}\eta^2) \;,
\label{D.11}\\[2mm]
&& \frac{\dr  G}{\dr \eta}  
= \lb \lp \tfrac{9}{2} \eta G - \tfrac{3}{2} H \rp
+ v \lp \tfrac{9}{4} \eta X - \tfrac{3}{2} Y - \tfrac{1}{2}  Z \rp \rb 
/(1-\tfrac{27}{4}\eta^2) \;,
\label{D.12}\\[2mm]
&& \frac{\dr H}{\dr \eta}  
= \lb \lp - G + \tfrac{9}{4} \eta H \rp
+ v \lp - \tfrac{1}{2} X + \tfrac{9}{4} \eta Y + \tfrac{3}{4} \eta Z \rp \rb
/(1-\tfrac{27}{4}\eta^2) \;,
\label{D.13}\\[2mm]
&& \frac{\dr X}{\dr \eta}  
= \lb \lp \tfrac{9}{4}  \eta X - Z \rp
+ v \lp - \tfrac{9}{4}  \eta F - \tfrac{3}{4} \eta G + \tfrac{1}{2} H \rp \rb 
/(1-\tfrac{27}{4}\eta^2) \;,
\label{D.14}\\[2mm]
&& \frac{\dr Y}{\dr \eta}  
=  \lb \lp X - 3\, \eta Z \rp 
+v \lp - F - \tfrac{9}{4} \eta^2 G + \tfrac{3}{2} \eta H \rp \rb 
/(1-\tfrac{27}{4}\eta^2) \;,
\label{D.15}\\[2mm]
&& \frac{\dr Z}{\dr \eta}  
=  \lb \lp -\tfrac{3}{2} X + \tfrac{9}{2}  \eta Z \rp 
+ v \lp \tfrac{3}{2}  F + \tfrac{1}{2} G - \tfrac{9}{4} \eta H \rp \rb 
/(1-\tfrac{27}{4}\eta^2) \;.
\label{D.16}
\eea

\subsection{derivatives of vectors}

Various combinations of the unit vectors $\hat{v}$ and $\hat{b}$ are also needed
and they are listed below for convenience.
Results from App.\ref{auxfunc} yield 
\bea
&& \hat{v}_i \, \hat{v}_j \, d_{ijk} = \tfrac{1}{\sqrt{3}} \, \hat{b}_k \;,
\label{D.17}\\
&& \hat{v}_i \, \hat{b}_j \, d_{ijk} = \tfrac{1}{\sqrt{3}} \, \hat{v}_k \;,
\label{D.18}\\
&& \hat{b}_i \, \hat{b}_j \, d_{ijk} = 3\, \eta \, \hat{v}_k - \tfrac{1}{\sqrt{3}} \, \hat{b}_k \;.
\label{D.19}
\eea
For terms involving derivatives, one uses $\dr v/\dr v_i = \hat{v}_i$ and finds
\bea
&& \frac{\dr \hat{v}_s}{\dr v_a} 
= \frac{1}{v}\lp \d_{as} - \hat{v}_a \, \hat{v}_s \rp \;,
\label{D.20}\\[2mm]
&& \frac{\dr \hat{b}_s}{\dr v_a}  
= \frac{2\sqrt{3}}{v} \lp \hat{v}_j \, d_{jas} - \tfrac{1}{\sqrt{3}}\, \hat{v}_a \, \hat{b}_s\rp \;,
\label{D.21}\\[2mm]
&& \hat{v}_s \,  \frac{\dr \hat{v}_s}{\dr v_a} =0 \;,
\label{D.22}\\[2mm]
&& \hat{b}_s \, \frac{\dr \hat{v}_s}{\dr v_a}  
=  \frac{1}{v}\lp - \hat{\bv}\cdot\hat{\bb} \;\hat{v}_a + \hat{b}_a \rp \;,
\label{D.23}\\[2mm]
&& \hat{v}_s \,  \frac{\dr \hat{b}_s}{\dr v_a}  
= \frac{2}{v} \lp - \hat{\bv}\cdot\hat{\bb} \; \hat{v}_a + \hat{b}_a \rp \;,
\label{D.24}\\[2mm]
&& \hat{b}_s \, \frac{\dr \hat{b}_s}{\dr v_a} =0 \;,
\label{D.25}\\[2mm]
&& \hat{v}_j \, \frac{\dr \hat{v}_s}{\dr v_a} \, d_{jsk} 
= \frac{1}{v}\lp \hat{v}_j \, d_{ajk} - \tfrac{1}{\sqrt{3}} \, \hat{v}_a \, \hat{b}_k \rp  \;,
\label{D.26}\\[2mm]
&& \hat{b}_j \, \frac{\dr \hat{v}_s}{\dr v_a}  \, d_{jsk}
= \frac{1}{v}\lp - \tfrac{1}{\sqrt{3}} \, \hat{v}_a\,\hat{v}_k + \hat{b}_j\, d_{ajk} \rp \;,
\label{D.27}\\[2mm]
&& \hat{v}_j \, \frac{\dr \hat{b}_s}{\dr v_a}  \, d_{jsk}
= \frac{1}{v} \lp \tfrac{1}{\sqrt{3}} \, \d_{ak} - \hat{b}_j\, d_{ajk} \rp \;,
\label{D.28}\\[2mm]
&& \hat{b}_j \, \frac{\dr \hat{b}_s}{\dr v_a} \, d_{jsk}
= \frac{1}{v} \lp \tfrac{3}{2}\, \eta \, \d_{ak} - 6\, \eta \, \hat{v}_a\, \hat{v}_k
- \hat{v}_j \, d_{ajk}  + \tfrac{1}{\sqrt{3}}\, \hat{v}_a \, \hat{b}_k 
+ \tfrac{3}{\sqrt{3}} \, \hat{b}_a \, \hat{v}_k \rp \;.
\label{D.29}
\eea
This allows one to write
\bea
&& \frac{\dr \eta}{\dr v_a} 
= \frac{2}{3\,\sqrt{3}} \, \frac{\dr \, (\hat{v}\cdot\hat{b})}{\dr v_a}
= \frac{1}{v} \lp - 3\, \eta\, \hat{v}_a  + \frac{2}{\sqrt{3}} \, \hat{b}_a \rp \;,
\label{D.30}\\[2mm]
&& \hat{b}_s \, \frac{\dr \hat{v}_s}{\dr v_a}  
= \frac{\sqrt{3}}{2}\, \frac{\dr \eta}{\dr v_a} \;,
\label{D.31}\\[2mm]
&& \hat{v}_s \,  \frac{\dr \hat{b}_s}{\dr v_a}  
= \sqrt{3}\; \frac{\dr \eta}{\dr v_a} \;,
\label{D.32}
\eea
and
\bea
&& \dr^\m v_a = \lp \dr^\m v \rp \, \hat{v}_a + v \, \dr^\m \hat{v}_a \;,
\label{D.33}\\[2mm]
&& \frac{\dr \hat{v}_s}{\dr v_a} \, \dr^\m v_a  = \dr^\m \hat{v}_s \;,
\label{D.34}\\[2mm]
&& \frac{\dr \hat{b}_s}{\dr v_a} \, \dr^\m v_a   = \dr^\m \hat{b}_s 
\label{D.35}\\[2mm]
&& \frac{\dr \eta}{\dr v_a} \,\dr^\m v_a = \dr^\m \eta
\label{D.36}
\eea

\subsection{results}

Recalling eqs.(\ref{anf.55}), (\ref{anf.56}), and using $\dr/\dr v_a \rar \dr_a$,
$\dr/\dr v \rar \dr_v$ and $\dr/\dr \eta \rar \dr_\eta$,
one writes
\bea
&& \hspace{-5mm}
\lb S\, \dr^\m W - W\, \dr^\m S + Q_i \, \dr^\m P_i - P_i \, \dr^\m Q_i \rb
\nn\\[2mm]
% && = 
% \lb F \, \dr_a \lp Y +\tfrac{2}{3} Z \rp
% + G \, \dr_a \lp \tfrac{2}{3} Y + \tfrac{2}{3} Z  + \eta  X \rp
% + H \, \dr_a \lp \tfrac{2}{3}  X +\eta Z \rp
% \right.
%
% \nn\\[1mm]
% && \left.
% - \, Y \, \dr_a \lp F +\tfrac{2}{3} G \rp
% - Z\, \dr_a  \lp \tfrac{2}{3} F + \tfrac{2}{3} G + \eta H \rp
% - X\,  \dr_a \lp \tfrac{2}{3} H  +\eta G \rp \rb 
%
% \nn\\[1mm]
% && 
% + \tfrac{2}{3\sqrt{3}} \lp H Z - G X \rp 
% \lp \hat{v}_s\, \dr_a \hat{b}_s   - \hat{b}_s\, \dr_a \hat{v}_s  \rp  
% \nn\\[3mm]
%
%
&& = 
\lb F \, \dr_v \lp Y  +\tfrac{2}{3} Z  \rp
+ G \, \dr_v \lp \tfrac{2}{3} Y + \tfrac{2}{3} Z + \eta X \rp
+ H \, \dr_v \lp \tfrac{2}{3} X +\eta Z \rp
\right.
\nn\\[1mm]
&& \left.
- \, Y \, \dr_v \lp F  +\tfrac{2}{3} G \rp
- Z \, \dr_v \lp \tfrac{2}{3} F + \tfrac{2}{3} G + \eta H \rp
- X \, \dr_v \lp \tfrac{2}{3}  H +\eta G \rp \rb \hat{v}_a
\nn\\[1mm]
&& 
+ \, \lb F \, \dr_\eta \lp Y +\tfrac{2}{3} Z  \rp
+ G \, \dr_\eta \lp \tfrac{2}{3} Y  + \tfrac{2}{3} Z + \eta X \rp
+ H \, \dr_\eta \lp \tfrac{2}{3} X +\eta Z \rp
\right.
\nn\\[1mm]
&& \left.
- \, Y \, \dr_\eta \lp F +\tfrac{2}{3} G \rp
- Z \, \dr_\eta \lp \tfrac{2}{3} F + \tfrac{2}{3} G  + \eta H \rp
- X \, \dr_\eta \lp \tfrac{2}{3} H  +\eta G  \rp
\right.
\nn\\[1mm]
&& \left.
+ \tfrac{1}{3} \lp H Z - G X \rp \rb \dr_a \eta 
\nn\\[3mm]
&& = \lb \eta\, C_B + \tfrac{2}{3}  C_A \rb \hat{v}_a
+ \lb \tfrac{1}{3} \, v \, C_B \rb \dr _a\eta  = 0 \;,
\label{D.37}
\eea
after transforming terms involving derivatives  into binomials of the basic functions 
by means of   eqs.(\ref{anf.23}), (\ref{anf.24}),  (\ref{D.11})-(\ref{D.16}),
and using results from App.\ref{unitproof}.

The vector contribution is obtained by a straightforward calculation and reads 
\bea 
&&  \hspace{-5mm}
 i \,  \tfrac{3}{2} \,  \lp Q_i\, \dr^\m Q_j  + P_i\, \dr^\m  P_j  \rp  f_{ijk} 
= i\, \lb \lp H^2 + X^2 \rp  \hat{v}_i \, \dr^\m \hat{v}_j 
\right.
\nn\\[2mm]
&& \left. + \, \tfrac{1}{\sqrt{3}}  \lp G H + X Z \rp 
\lp  \hat{v}_i \, \dr^\m \hat{b}_j + \hat{b}_i \, \dr^\m \hat{v}_j \rp 
+  \tfrac{1}{3} \lp G^2 + Z^2 \rp \hat{b}_i \, \dr^\m \hat{b}_j  \rb  f_{ijk}  \;.
\label{D.38}
\eea 

The axial term is 
\bea
&& \hspace{-5mm} 
i\,  \lb \sqrt{\tfrac{3}{2}} \lp S\, \dr^\m P_k  - P_k\, \dr^\m S 
- W\, \dr^\m Q_k + Q_k \, \dr^\m W   \rp 
+ \tfrac{3}{2} \lp  Q_i\, \dr^\m P_j 
- P_i\, \dr^\m Q_j \rp d_{ijk} \rb
\nn \\[2mm]
&& = i\, \lc \lb  \lp F+G \rp \dr_a X 
+ H \, \dr_a Y 
+ \lp H + \eta G\rp \dr_a Z 
- \lp Y + Z \rp  \dr_a H  
- X\, \dr_a F 
- \lp X + \eta Z \rp  \dr_a G  \rb \hat{v_k} 
\right.
\nn\\[2mm]
&& \left.
+ \, \tfrac{1}{\sqrt{3}} \lb  \lp F + G \rp \dr_a Z  
+ G\, \dr_a Y  + H \, \dr_a X 
-  \lp Y + Z \rp \dr_a G 
- Z\,  \dr_a F - X \, \dr_a H  \rb  \hat{b}_k
\right.
\nn\\[2mm]
&& \left. 
+ \lb \lp F + \tfrac{2}{3} G \rp  X - \lp Y + \tfrac{2}{3}  Z \rp H  \rb \dr_a \hat{v}_k 
+  \tfrac{1}{\sqrt{3}} \lb \lp F +\tfrac{2}{3}  G \rp  Z - \lp Y + \tfrac{2}{3} Z \rp G \rb 
\dr_a \hat{b}_k
\right.
\nn\\[2mm]
&& \left. + \tfrac{1}{\sqrt{3}} \, \lp  H Z - G X \rp 
\lp \hat{v}_i \, \dr_a \hat{b}_j - \hat{b}_i \, \dr_a \hat{v}_j \rp  d_{ijk}
\rc \dr^\m v_a
\label{D.39}
\eea
Reexpressing  the derivatives by means of (\ref{D.2}) and employing
results from App.\ref{unitproof}, one has 
\bea
&& \hspace{-5mm} 
i\,  \lb \sqrt{\tfrac{3}{2}} \lp S\, \dr^\m P_k  - P_k\, \dr^\m S 
- W\, \dr^\m Q_k + Q_k \, \dr^\m W   \rp 
+ \tfrac{3}{2} \lp  Q_i\, \dr^\m P_j 
- P_i\, \dr^\m Q_j \rp d_{ijk} \rb
\nn \\[2mm]
&& = i\, \lc \lb 1 \rb \hat{v}_k \, \dr^\m v \,
\right.
\nn\\[2mm]
&& \left.
+ \, \frac{1}{(1-\tfrac{27}{4}\eta^2)} 
\lb  \lp G Y - F Z \rp  + \tfrac{9}{4} \eta \lp F X - H Y \rp 
+ \tfrac{3}{4} \eta \lp HZ-G X \rp - \tfrac{9}{4}\, v\, \eta  
\rb  \hat{v_k}  \, \dr^\m  \eta 
\right.
\nn\\[2mm]
%
%4
&& \left.
+ \, \frac{1}{ (1-\tfrac{27}{4}\eta^2)} \, \tfrac{1}{\sqrt{3}}
\lb  \tfrac{3}{2} \lp H Y - F X \rp  + \tfrac{1}{2} \lp G X - H Z \rp 
+ \tfrac{9}{2} \eta \lp F Z - G Y \rp 
+  \tfrac{3}{2} \,  v \rb  \hat{b}_k \, \dr^\m \eta 
\right.
\nn\\[2mm]
%
%5
&& \left. 
+ \lb \lp F + \tfrac{2}{3} G \rp  X - H \lp Y + \tfrac{2}{3}  Z \rp  \rb 
\dr^\m \hat{v}_k 
+  \tfrac{1}{\sqrt{3}}  \lb \lp F +\tfrac{2}{3}  G \rp  Z - G \lp Y + \tfrac{2}{3} Z \rp \rb 
\dr^\m \hat{b}_k
\right.
\nn\\[2mm]
&& \left. 
+ \tfrac{1}{\sqrt{3}}  \lp HZ - G X \rp 
\lp \hat{v}_i \, \dr^\m \hat{b}_j - \hat{b}_i \, \dr^\m \hat{v}_j \rp  d_{ijk} \rc 
\label{D.40}
\eea

% REFERENCES ----------------------------------------------------------------------


\begin{thebibliography}{99}



\bibitem{LattQCD}  J. Kogut and L. Susskind, Phys. Rev. D 11, 395 (1975);
L. Susskind, Phys. Rev. D16 3031 (1977);
H. Hamber and G. Parisi, Phys. Rev. Lett. 47 1792 (1981);
E. Marinari, G. Parisi and C. Rebbi, Phys. Rev. Lett. 47 1795 (1981);
R. A. Briceno, J. J. Dudek and R. D. Young, 
Rev. Mod. Phys. 90, 025001 (2018).

\bibitem{GellMannLevy} M. Gell-Mann and M. Levy, Nuovo Cimento 16, 705 (1960).

\bibitem{Schwinger}  J. Schwinger, Ann. Phys., 2, 407 (1957).

\bibitem{NambuJonaL} Y. Nambu and G. Jona Lasinio, 
Phys. Rev. 124, 246 (1961);
Phys. Rev. Lett. 4, 380 (1960); 
Phys. Rev. 122, 345 (1961).


\bibitem{Skyrme} T. H. R. Skyrme, Proc. Roy. Soc. Lond. A 260,  127 (1961);
Nucl. Phys. 31, 556 (1962);
Proc. R. Soc. Lond. A 247, 260 (1958);
for the origins of skyrmions,  see T. H. R. Skyrme, Int. J. Mod. Phys. A 3,  2745 (1988);
I. J. R. Aitchison, arXiv:2001.09944v2.


\bibitem{AdNaWit} G. S. Adkins, C. R. Nappi and E. Witten E Nucl. Phys. B 228, 552 (1983).


\bibitem{Weinberg1} S. Weinberg, Physica A 96, 327 (1979);
see also arXiv:hep-th/0908.1964v3, for reminiscences of that period

\bibitem{GassLeut1}  J. Gasser and H. Leutwyler, Ann. Phys. 158, 142 (1984).

\bibitem{Levy}  M. L\'evy, Nuovo Cimento 52, 23 (1967).

\bibitem{GassLeut2}
J. Gasser and H. Leutwyler, Nucl. Phys. B250, 465 (1985).

\bibitem{Chemtob} M. Chemtob, N. Cim. 89 A, 381 (1995)

\bibitem{ChemtobBlaizot} M. Chemtob, N. Phys. B 256, 600 (1985);
M. Chemtob and J. P. Blaizot, N. Phys. A 451, 605 (1986).

\bibitem{Nelson} T. J. Nelson, J. Math. Phys. 8, 957 (1967).

\bibitem{AkyRas} D. A. Akyeampong and M. A. Rashid, J. Math. Phys. 13, 1218 (1972).

\bibitem{Byrd} M. Byrd, J. Math. Phys. 39, 6125 (1998); 
Erratum, J. Math. Phys. 41, 1026 (2000);
M. Byrd and E. C. G. Sudarshan, J.Phys.A 31, 9255 (1998).

\bibitem{color} P. Amore, M. C. Birse, J. A. McGovern, N. R. Walet, 
Phys.Rev.D 65, 074005 (2002);  
Int.J.Mod.Phys.B 17, 5185 (2003), 
% Ser. Adv. Quant. Many Body Theor. 6, 226 (2002), 
% Acta Phys. Hung. A 16,  163 (2002).

\bibitem{optics}  S. Asthana and , V. Ravishankar,  J. Opt. Soc. Am. B 39, 1 (2022);
 J. Opt. Soc. Am. B, 433075 (2021);
C. M. Caves and G. J. Milburn, Opt. Comm. 179, 439 (2000).


\bibitem{geompha} Z. S. Wang et al. Physical Review A 75, 024102 (2007);
E. Ercolessi et al., Int. J.  Mod. Phys. A 16, 5007 (2001).


\bibitem{entanglement} K. Zyczkowski and H-J. Sommers,  J. Phys A 36, 10115 (2003);
E. Sj\"oqvist, Phys. Rev. A 62, 022109 (2000);
P. B. Slater, J. Physics A 32, 5261 (1999);
P. B. Slater, Eur. Physl J. B, 471 (2000);


\bibitem{Weinberg68} S. Weinberg, Phys. Rev. 166, 1568 (1968).

\bibitem{Gasio} S. Gasiorowicz, Elementary Particle Physics, John Wiley and Sons,
New York, London and Sydney, 1967


\end{thebibliography}
\end{document}